\documentclass[10pt,aps,prd,notitlepage,preprintnumbers]{revtex4-1}
\usepackage{graphicx}
\usepackage{slashed}
\usepackage{xcolor}
\usepackage{amsmath}
\usepackage{amssymb}
\usepackage{amsfonts}
\usepackage[bookmarks=false,colorlinks, linkcolor=green]{hyperref}
\newcommand{\ep}{\epsilon}
\newcommand{\la}{\lambda}
\newcommand{\de}{\delta}
\newcommand{\ga}{\gamma}
\newcommand{\al}{\alpha}
\newcommand{\be}{\beta}
\newcommand{\sig}{\sigma}
\newcommand{\s}[1]{\slashed{#1}}

\newcommand{\no}{\notag\\}

\begin{abstract}
It is known the single transverse spin asymmetry in semi-inclusive deep inelastic scattering can be factorized by a twist-3 distribution function $T_F$, which contains a gluon field strength tensor. With transverse gluon included in the power expansion, we find the gluon field strength tensor can be recovered definitely in soft-gluon-pole contribution at leading order of $\al_s$ expansion. This conclusion holds in Feynman and light-cone gauges.
\end{abstract}

\begin{document}
\title{Role of transverse gluon in SSA}
\author{G.P. Zhang}
\email[]{gpzhang@ynu.edu.cn}
\affiliation{Department of physics, Yunnan University, Kunming, Yunnan 650091, China}
\maketitle

\section{Introduction}
The single transverse spin asymmetry(SSA) in high-$P_T$ pion production is discovered in 70's\cite{Klem:1976ui}, and the asymmetry is of order $10\%$.
A possible explanation in perturbative QCD for this large asymmetry is
Efremov-Teryaev-Qiu-Sterman(ETQS) mechanism, which has been proposed
for many years\cite{Efremov:1981sh,*Efremov:1984ip},\cite{Qiu:1991pp,*Qiu:1991wg}. In this mechanism, SSA is proportional to ETQS matrix
element, which is a correlation function of quark and gluon fields defined on
light-cone. How to obtain the twist-3 hard coefficients before ETQS matrix element
has been discussed thoroughly in literatures. One of the remaining problems is
how to recover gluon field strength tensor appearing in ETQS matrix element consistently. A clear algorithm to get the twist-3 hard coefficients is first given by Qiu and Sterman\cite{Qiu:1991pp,*Qiu:1991wg}, in which the subcross section is calculated by taking the initial coherent gluon as a longitudinal gluon $G^+$ with a small transverse momentum $k_\perp$, then,
the subcross section is expanded to $O(k_\perp)$. After the proof that
only the matrix element $\langle \bar{\psi}(\partial_\perp^\rho G^+)\psi\rangle$ appears after transverse momentum
expansion, the ETQS matrix element $T_F$ then
is obtained by the replacement $\partial_\perp^\rho G^+\rightarrow -G_\perp^{+\rho}$, where $G_\perp^{+\rho}$ is gluon field strength tensor. This
replacement is applied in almost all following works, see for example \cite{Qiu:1998ia,Kanazawa:2000cx,Kouvaris:2006zy,Ji:2006vf,Ji:2006br,Eguchi:2006qz}. In \cite{Vogelsang:2009pj,Chen:2016dnp,*Chen:2017lvx,Kang:2012ns,Dai:2014ala,
Yoshida:2016tfh,Benic:2019zvg}even loop
correction is calculated in this formalism, and the twist-3 factorization formula
is justified at one-loop level. Thus, this algorithms is reliable to give correct
answer. But, since the contribution of $G_\perp$ is not calculated, there is still a problem whether the contribution of $G_\perp$ can be incorporated into ETQS matrix element. For a complete calculation, one has to also calculate the contribution of $G_\perp$ and to see whether the combination $\partial_\perp^\rho G^+ - \partial^+ G_\perp^\rho$ appears.
This problem is studied by Eguchi, Koike and Tanaka in\cite{Eguchi:2006mc},
where a group of consistence relations are derived in order to make sure gluon
field strength tensor $G_\perp^{+\rho}$ is correctly(completely) reproduced. For hard-gluon-pole and soft-fermion-pole contributions, these conditions are satisfied easily due to Ward Identities, and it is confirmed that $G_\perp$ expansion gives the same hard coefficients as that obtained from $G^+$ expansion.
However, for soft-gluon-pole(SGP) contribution, although the conditions are
satisfied by analyzing the detailed cancellation between mirror diagrams, a
direct calculation based on $G_\perp$ expansion is still missing.  For this reason how to obtain SGP in light-cone gauge is not described in \cite{Eguchi:2006mc} either. It is argued that
$G_\perp$ contribution contains some ambiguities and some hard coefficients may
be lost due to $x\de(x)=0$. The analysis of \cite{Eguchi:2006mc} is very clear and thorough. But, as we will show in this paper, $G_\perp$ expansion is definite and gives the same hard coefficients as that obtained from $G^+$ expansion. The price for using $G_\perp$ expansion is one has to incorporate
the contribution from another twist-3 distribution function $q_\partial(x)$,
besides ETQS matrix element $T_F(x,x)$.
Also, the coefficient of $q_\partial(x)$ can be determined definitely.
As an example, we will consider the SSA for high $P_T$ pion production in semi-inclusive deep inelastic scattering(SIDIS), which is considered in \cite{Ji:2006br},\cite{Eguchi:2006qz},\cite{Eguchi:2006mc}. The generalization of this proof to other processes is not difficult. The paper is organized as follows: In Sec.II, our notations and the kinematics of SIDIS are introduced;
In Sec.III, the calculation including $G_\perp$ expansion is performed and
how to get the gluon fields strength tensor $G_\perp^{+\rho}$ is shown explicitly,
and a formula is given for the corresponding hard coefficients;
In Sec.IV, the explicit expressions of hard coefficients for quark and gluon fragmentations are given;
In Sec.V, we shortly discuss the generalization of our proof to higher orders of $\al_s$ expansion and make a summary.

\section{Notations and kinematics}
We work with light-cone coordinates throughout this paper, for which the components
of an arbitrary vector $a^\mu$ are $(a^+,a^-,a_\perp^\mu)$. The transverse direction
is defined by two light-like vectors $l^\mu,n^\mu$, and the transverse metric is
\begin{align}
g_\perp^{\mu\nu}=g^{\mu\nu}-\frac{l^\mu n^\nu+l^\mu n^\nu}{n\cdot l}.
\end{align}
Then, $a^\pm$ are defined as $a^+=n\cdot a$, $a^-=l\cdot a$ and the transverse
component is $a_\perp^\mu=g_\perp^{\mu\nu}a_\nu$.

For a hadron(proton) moving along $Z-$axis with a large $p_A^+$,
twist-3 distribution functions are defined as
\begin{align}
&\int \frac{d\xi^-d\xi_1^-}{(2\pi)^2}e^{i\xi^- xP^+ +i\xi_1^- x_1P^+}
\langle Ps_\perp|
\bar{\psi}_j(0)[g_sG_{a\perp}^{+\rho}(\xi^-)\psi(\xi_1^-)]_i|Ps_\perp\rangle
=\frac{\Big[\ga^- T^a\tilde{s}_\perp^\rho T_F(x_1,x_2) +i\ga_5\ga^-T^a s_\perp^\rho T_{\Delta F}(x_1,x_2)\Big]_{ij}}{4\pi N_cC_F},\no
&\int \frac{d\xi^-}{2\pi}e^{i\xi^- xP^+}
\langle Ps_\perp|
[\bar{\psi}(0)\mathcal{L}_n^\dagger(0)]_j i\partial_\perp^\rho [\mathcal{L}_n(\xi^-)\psi(\xi_1^-)]_i|Ps_\perp\rangle
=\frac{1}{2N_c}\Big[\ga^-\tilde{s}_\perp^\rho q_\partial(x)
+\ga_5\ga^-s_\perp^\mu\tilde{q}_\partial(x)\Big]_{ij},
\label{eq:T_F}
\end{align}
with $x_2=x_1+x$ and $\tilde{s}_\perp^\rho=\ep^{-+\rho\tau}s_{\perp\tau}$.

For a hadron(proton) moving along $-Z-$axis with a large $p_H^-$, the fragmentation functions for quark and gluon are\cite{Collins:1981uw}
\begin{align}
\frac{1}{z}D_{q}(z)=&\int\frac{d\xi^+}{2\pi}
e^{-ik^- \xi^+}\langle 0|\psi_i(0)|p_HX\rangle \langle Xp_H|\bar{\psi}_j(\xi^+)|0\rangle,\no
\frac{1}{z}D_{g}(z)=&\frac{-1}{(N_c^2-1)2k^-}\int
\frac{d\xi^+}{2\pi} e^{-i k^- \xi^+}
\langle 0|G_{b}(0)^{-\al}|p_HX\rangle\langle p_HX|G_{b}(\xi^+,0,0_\perp)^{-\be}g_{\perp\al\be}
|0\rangle,\ k^-=1/z p_H^-.
\end{align}
The gauge link is defined as
\begin{align}
\mathcal{L}_n(\xi^-)=Pe^{-ig_s\int_0^\infty d\la^- G^+(\xi^-+\la^-)}.
\label{eq:gauge_link}
\end{align}
For simplicity, the gauge link is suppressed in the above definitions, if there is no derivative acting on the gauge link. In Feynman gauge, if there is no gauge link, from time-reversal and parity symmetries one can show very easily that $q_\partial=0$. In this calculation, the covariant derivative is $D^\mu=\partial^\mu +ig_sT^aG_a^\mu$, and the anti-symmetric tensor satisfies $\ep^{0123}=1$.

As an example, in this work we consider the SSA for pion production in SIDIS.
The process is
\begin{align}
e(l_e)+h_A(p_A,s_{a\perp})\rightarrow e(l'_e)+\pi(p_H)+X,
\end{align}
where $X$ represents undetected hadrons. The momenta and spin of particles are
written in the brackets. Further, in our case the exchanged vector boson is
virtual photon only. For kinematics, we mainly use the notations in \cite{Eguchi:2006mc}.
We work in hadron frame, where initial hadron and virtual photon are moving along $+Z$ and $-Z$-axis, respectively. In this frame we demand the final hadron
has a large transverse momentum with respect to $Z-$axis, i.e.,  $p_{H\perp}\gg\Lambda_{QCD}$. With respect to the hadron plane expanded by
$p_A$ and $p_H$, the azimuthal angle of final lepton is $\phi$ and the azimuthal
angle of spin vector $\vec{s}_{a\perp}$ is $\Phi_s$. Other invariants are standard:
\begin{align}
x_B=\frac{Q^2}{2p_A\cdot q},\ z_f=\frac{p_A\cdot p_H}{p_A\cdot q},\
y=\frac{p_A\cdot q}{p_A\cdot l_e},\ S_{ep}=(l_e+p_A)^2,\ Q^2=-q^2,\ q=l_e-l'_e.
\end{align}
To define $\pm$ components of momenta, we choose $p_A$ and $p_H$ as the two light-like vectors, with $p_A^-=p_H^+=0$. Then, virtual photon has a transverse momentum $q_\perp^\mu$, i.e.,
\begin{align}
q^\mu=\frac{q\cdot p_H}{p_A\cdot p_H}p_A^\mu
+\frac{q\cdot p_A}{p_A\cdot p_H}p_H^\mu+q_\perp^\mu.
\end{align}
Following differential cross section will be studied
\begin{align}
d\sig=&\frac{1}{2S_{ep}}\frac{d^3 p_H}{(2\pi)^3 2p_H^0}
\frac{d^3 l'}{(2\pi)^3 l^{'0}}\frac{e^4}{q^4}L^{\mu\nu}(2\pi)^4W_{\mu\nu}\no
=& dx_B dQ^2 d\phi dz_f dq_T^2\frac{\pi \al^2 z_f}{4S_{ep}^2 Q^2 x_B^2}
L_{\mu\nu}W^{\mu\nu},
\end{align}
where leptonic tensor is $L^{\mu\nu}=2(l_e^\mu l_e^{'\nu}+l_e^\nu l_e^{'\mu}
-g^{\mu\nu}Q^2)$, and hadronic tensor is
\begin{align}
W^{\mu\nu}=\int \frac{d^4x}{(2\pi)^4} e^{iq\cdot x}\sum_X\langle p_A,s_{a\perp}|
j^\nu(x)|p_H X\rangle\langle Xp_H|j^\mu(0)|p_A,s_{a\perp}\rangle,
\end{align}
with electro-magnetic current given by $j^\mu=\bar{\psi}\ga^\mu\psi$.

As done in\cite{Eguchi:2006mc}, one can introduce tensors $\mathcal{V}_k^{\mu\nu}$ to project out lepton azimuthal angle distributions. This gives
\begin{align}
L_{\mu\nu}W^{\mu\nu}=\sum_{k} A_k B_k,\
A_k=L_{\mu\nu}\mathcal{V}_k^{\mu\nu},\ B_k=W_{\mu\nu}\tilde{\mathcal{V}}^{\mu\nu}_k.
\label{eq:project_W}
\end{align}
All $\phi$ distributions are contained in $A_k$. It is found in \cite{Eguchi:2006mc} that only four distributions are relevant in SSA. They are
\begin{align}
A_1=Q^2[1+\cosh^2\psi],\ A_2=-2Q^2,\ A_3=-Q^2\cos\phi\sinh 2\psi,\
A_4=Q^2 \cos2\phi\sinh^2\psi,
\end{align}
where $\cosh\psi\equiv 2x_B S_{ep}/Q^2-1$.
$\mathcal{\tilde{V}}_k^{\mu\nu}$
are list in Appendix.\ref{sec:weights}. For more details, please consult
Ref.\cite{Eguchi:2006mc} and reference therein.

The typical hard scale of this process is $Q$, with $Q\gg \Lambda_{QCD}$.
$q_T\equiv\sqrt{-q_\perp^2}\sim Q$ is also taken as a hard scale. Since there is no
soft scale in $d\sig$, the differential cross section or hadronic tensor is
expected to be factorized in collinear formalism. In this paper we only consider the contribution of $q_\partial$ and $T_F$. Extending our calculation to include $\tilde{q}_\partial$ and $T_{\Delta F}$ is straightforward.

\section{SGP contribution}

\begin{figure}
\begin{center}
\begin{minipage}[b]{0.21\textwidth}
\includegraphics[width=\textwidth]{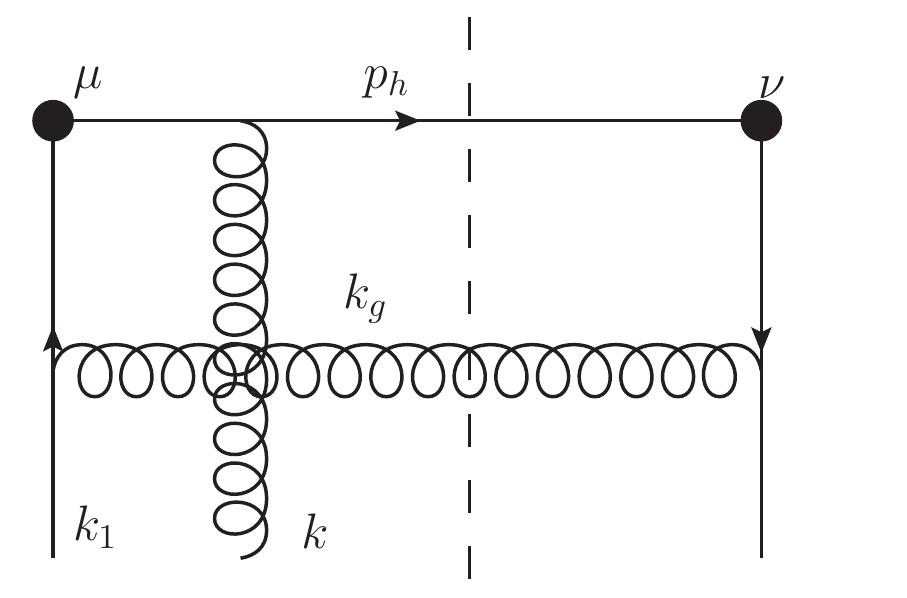}\\
(a)
\end{minipage}
\begin{minipage}[b]{0.21\textwidth}
\includegraphics[width=\textwidth]{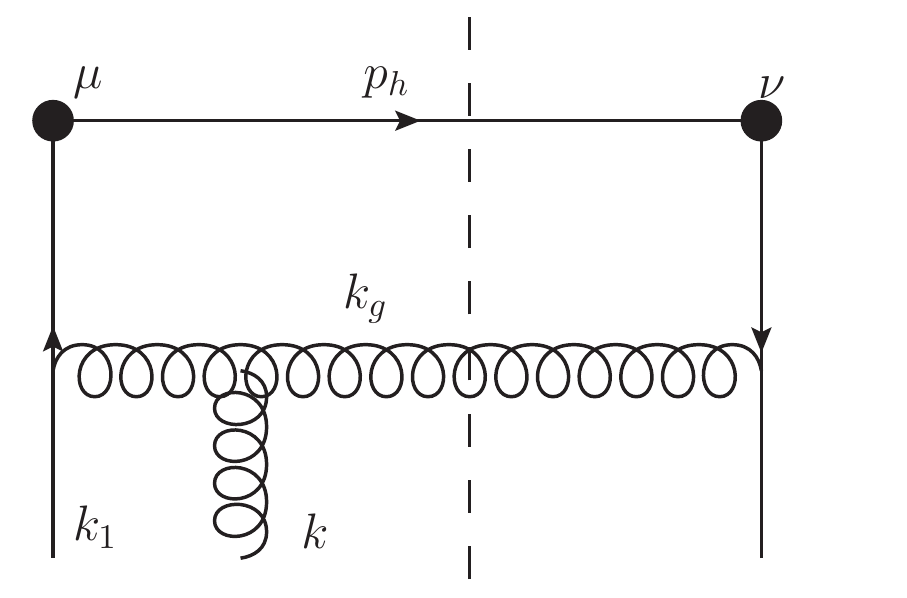}\\
(b)
\end{minipage}
\begin{minipage}[b]{0.2\textwidth}
\includegraphics[width=\textwidth]{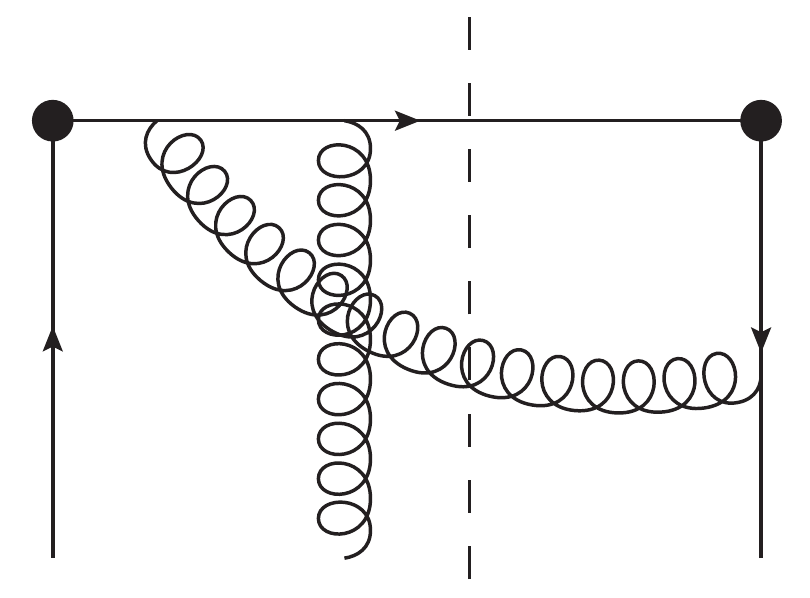}\\
(c)
\end{minipage}
\begin{minipage}[b]{0.2\textwidth}
\includegraphics[width=\textwidth]{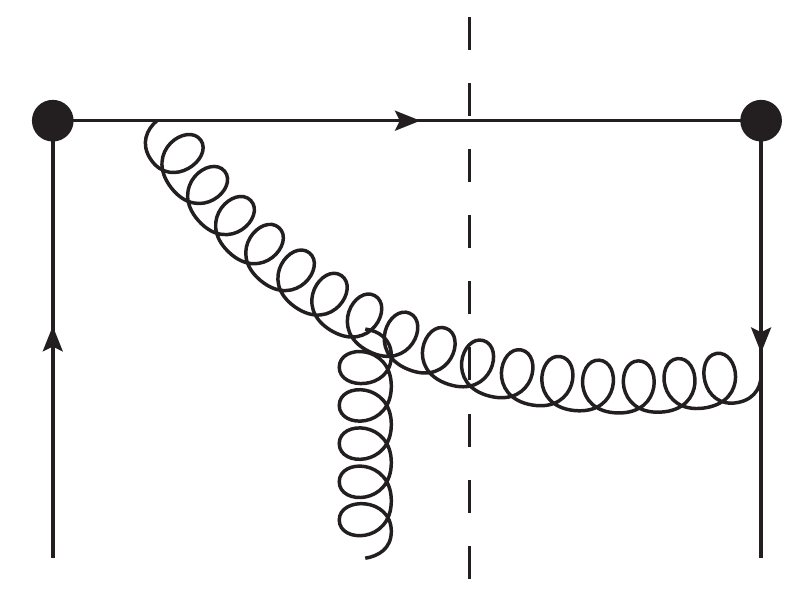}\\
(d)
\end{minipage}
\begin{minipage}[b]{0.2\textwidth}
\includegraphics[width=\textwidth]{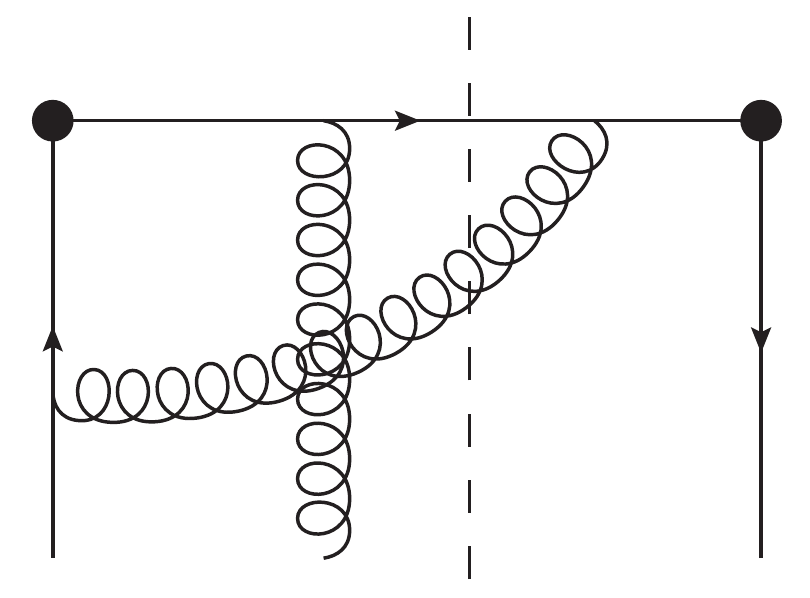}\\
(e)
\end{minipage}
\begin{minipage}[b]{0.2\textwidth}
\includegraphics[width=\textwidth]{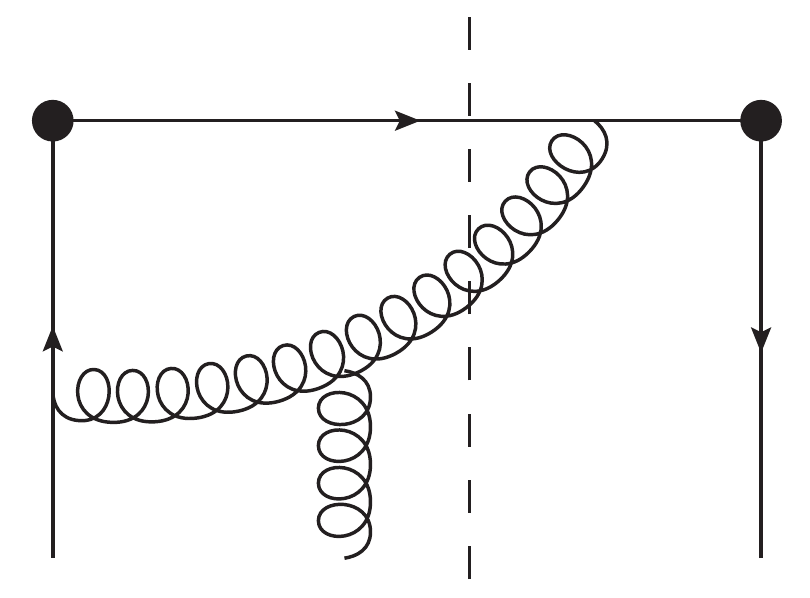}\\
(f)
\end{minipage}
\begin{minipage}[b]{0.2\textwidth}
\includegraphics[width=\textwidth]{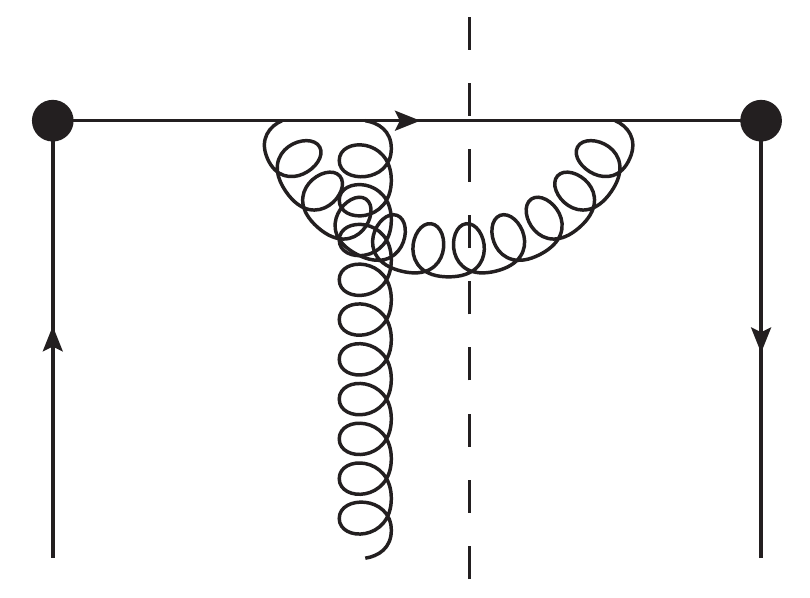}\\
(g)
\end{minipage}
\begin{minipage}[b]{0.2\textwidth}
\includegraphics[width=\textwidth]{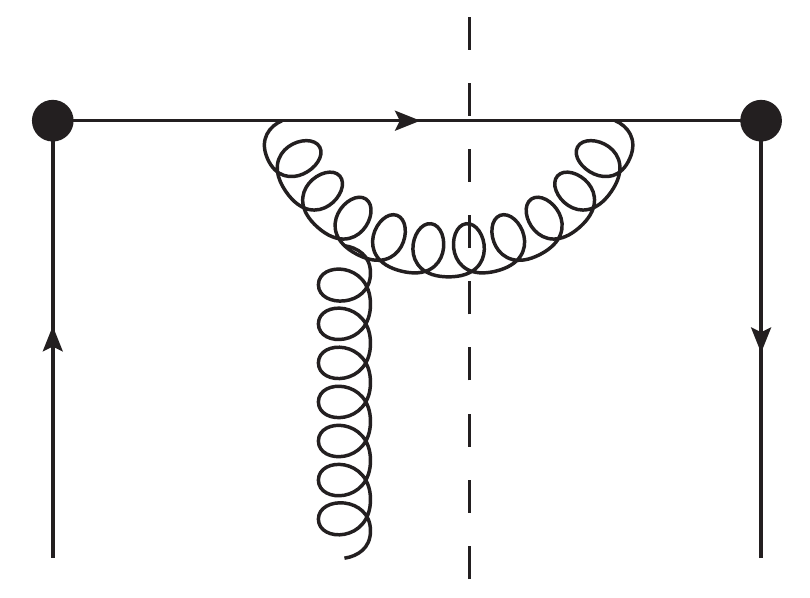}\\
(h)
\end{minipage}
\end{center}
\caption{Diagrams containing explicit SGP in SIDIS, where the dots are EM vertices.}
\label{fig:SSA}
\end{figure}

With the help of fragmentation function, the hadronic tensor can be written as
\cite{Eguchi:2006mc}
\begin{align}
W^{\mu\nu}(p_A,q,p_H)=& \sum_{j=q,g}\int \frac{dz}{z^2}D_j(z)
w_j^{\mu\nu}(p_A,q,p_h),\ p_h=\frac{1}{z}p_H,
\end{align}
where $D_j(z)$ is the fragmentation function for final hadron $h_B$ in parton $j$.
Here parton $j$ can be quark or gluon. Generally, $w^{\mu\nu}$ is
\begin{align}
w^{\mu\nu}=\int \frac{d^n k_1}{(2\pi)^4}\frac{d^nk}{(2\pi)^4}
H^{\mu\nu}_{a,\rho}(k_1,k)
\int d^n\xi_1 d^n \xi e^{ik\cdot \xi}e^{ik_1\cdot\xi_1}
\langle p_A,s_\perp|\bar{\psi}(0)g_s G_{a}^\rho(\xi)\psi(\xi_1)|p_A,s_\perp\rangle,
\end{align}
where $H^{\mu\nu}_{a,\rho}$ is the hard part. The main contribution is given by collinear region, where
\begin{align}
k^\mu=(k^+,k^-,k_\perp)\sim Q(1,\la^2,\la),\ k_1^\mu\sim Q(1,\la^2,\la).
\end{align}
Then we do power expansion in hard part $H^{\mu\nu\rho}_a$. At twist-3 level, $O(\la)$, both $G^+$ and $G_\perp$ contribute.
Consider the case with quark as fragmentation parton first.
In this case the momentum of final
gluon should be integrated. It is clear from previous studies that only pole
contributions give a real cross section\cite{Qiu:1991pp}. In this paper we just
consider the soft-gluon-pole(SGP) contribution.
The diagrams containing explicit SGP are shown in Fig.\ref{fig:SSA}.
The hard part of Fig.\ref{fig:SSA}(a) is
\begin{align}
H_{a,\rho}^{\mu\nu} g_sG_{a}^\rho(\xi^-)
=& \int \frac{d^n k_g}{(2\pi)^n}(2\pi)
\de(k_g^2)\de^n(k+k_1+q-p_h-k_g)P^{\be\be'}(k_g)\no
&\Big[(igT^b\ga_{\be'})\frac{-i}{\s{p}_h-\s{q}-i\ep}
\ga^\nu\s{p}_h(-igT^a\ga^\rho G_{a\rho}(\xi^-))[\frac{i}{\s{p}_h-\s{k}+i\ep}]
\ga^\mu\frac{i}{\s{p}_h-\s{k}-\s{q}+i\ep}(-igT^b\ga_\be)\Big],
\end{align}
where $P^{\be\be'}(k_g)$ is the tensor for the summation of polarization of final gluon. Firstly, we work in Feynman gauge $\partial_\mu G^\mu=0$.
For convenience we constrain final gluon to be physically polarized, then $P^{\be\be'}$ is chosen to be
\begin{align}
P^{\be\be'}(k_g)=-g^{\be\be'}+\frac{p_A^\be k_g^{\be'}+p_A^{\be'}k_g^\be}{
p_A\cdot k_g},
\end{align}
with $p_A$ the reference vector. In this way, $k_{g\be}P^{\be\be'}
=p_{A\be}P^{\be\be'}=0$.
The SGP is given by the propagator with momentum $p_h-k$. The expansion to twist-3 for following quantity is
\begin{align}
\s{p}_h(-igT^a\s{G}_a)\frac{i}{\s{p}_h-\s{k}+i\ep}
=&gT^a G_a^{+}\s{p}_h\Big[\frac{1}{-k^+ +i\ep}\Big]
+\frac{gT^a}{-2p_h^- k^+ +i\ep}\s{p}_h\ga^-
\ga_\perp^\rho(-k_{\perp\rho} G_a^+ + k^+ G_{a\perp\rho})
+O(\la^2).
\label{eq:a_expansion}
\end{align}
The first term is $O(1)$, and the second term is $O(\la)$, in which
gluon field strength tensor appears naturally. The SGP
contribution then is obtained by using following formula
\begin{align}
\frac{1}{-k^++i\ep}=P\frac{1}{-k^+}-i\pi \de(k^+).
\label{eq:pv}
\end{align}
Now the first term in eq.(\ref{eq:a_expansion}), which is $O(1)$,
gives $O(\la)$ contribution by expanding the
other parts of $H^{\mu\nu}_{a,\rho}$ in $k_\perp$ to $O(\la)$ or $O(k_\perp)$.
Here we ignore the
transverse momentum $k_{1\perp}$ first. We will come to $k_{1\perp}$ expansion later.
The expansion of $k_\perp$ contains two parts: one is from intermediate fermion
propagator $i/(\s{p}_h-\s{k}-\s{q})$, the other is from
$\de(k_g^2)P_{\be\be'}(k_g)$ for the final gluon, with $k_g=k_1+k+q-p_h$.
Expressed by diagrams, the expansion reads(for the rules of these diagrams, please
see Appendix.\ref{sec:rules})
\begin{align}
\text{Term-1}\equiv & \frac{G_a^+ k_\perp^\rho}{-k^+ +i\ep}
\frac{\partial}{\partial k_\perp^\rho}
\begin{matrix}
\includegraphics[scale=0.5]{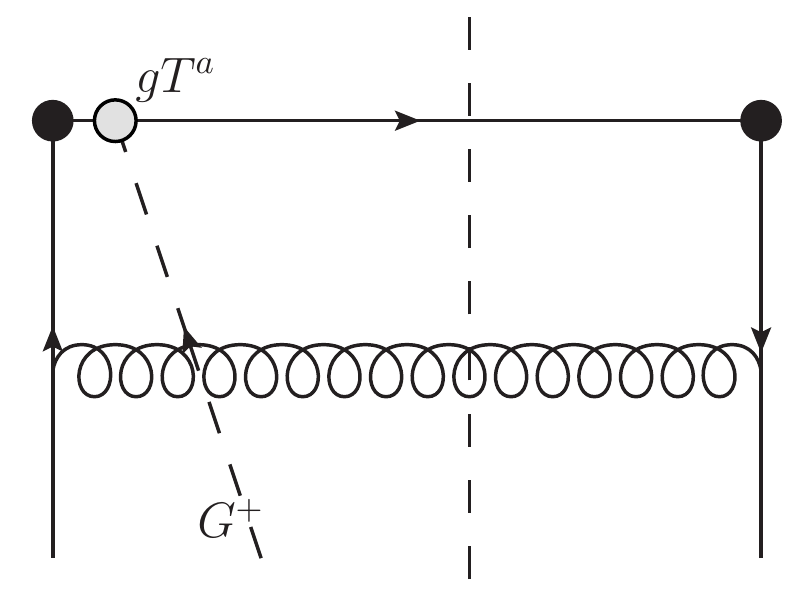}
\end{matrix},
\end{align}
and
\begin{align}
k_\perp^\rho\frac{\partial}{\partial k_\perp^\rho}
\begin{matrix}
\includegraphics[scale=0.5]{term-1}
\end{matrix}
=&
\begin{matrix}
\includegraphics[scale=0.5]{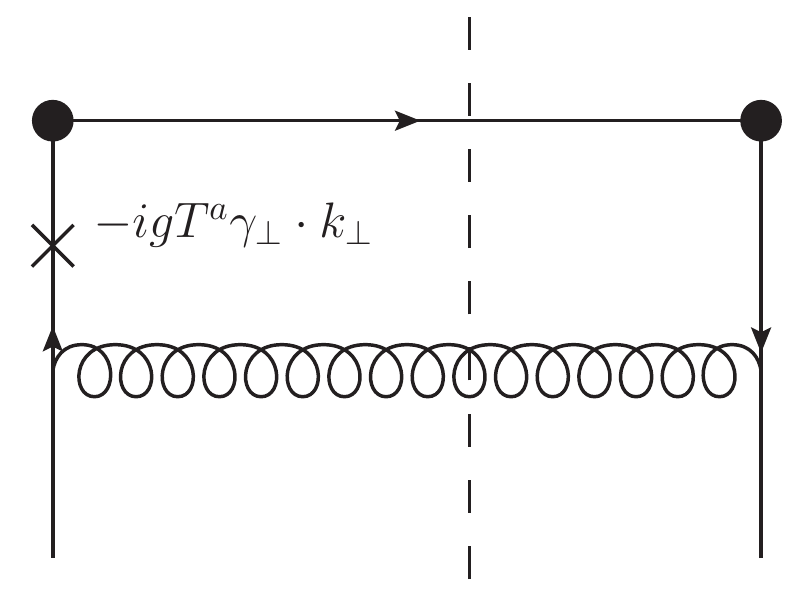}
\end{matrix}
+k_\perp^\rho\frac{\partial}{\partial k_\perp^\rho}
\begin{matrix}
\includegraphics[scale=0.5]{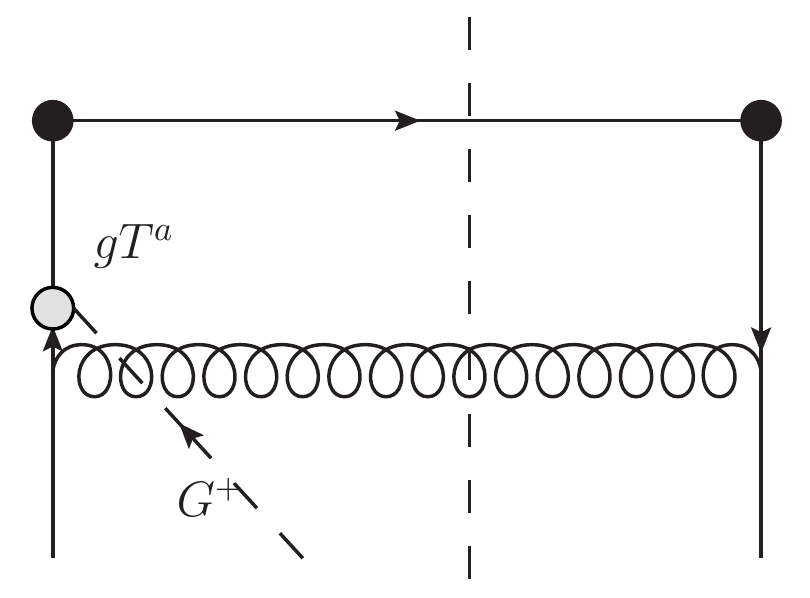}
\end{matrix}.
\label{eq:k_expansion}
\end{align}
For the expansion of fermion propagator, we have used the formula
\begin{align}
\frac{i}{\s{p}_h-\s{k}-\s{q}+i\ep}\doteq
\frac{i}{\s{p}_h-\s{q}+i\ep} (-i)\s{k}_\perp \frac{i}{\s{p}_h-\s{k}-\s{q}+i\ep}.
\end{align}
The expansion is equivalent to an insertion of transverse momentum.
Now, with the factor $gT^a$ included, the contribution is proportional to
$-igT^a \ga_{\perp\rho} k_\perp^\rho G_a^{+}$. This vertex is expressed in the
diagram as a cross on fermion propagator.

\begin{figure}
\begin{center}
\begin{minipage}[b]{0.2\textwidth}
\includegraphics[width=\textwidth]{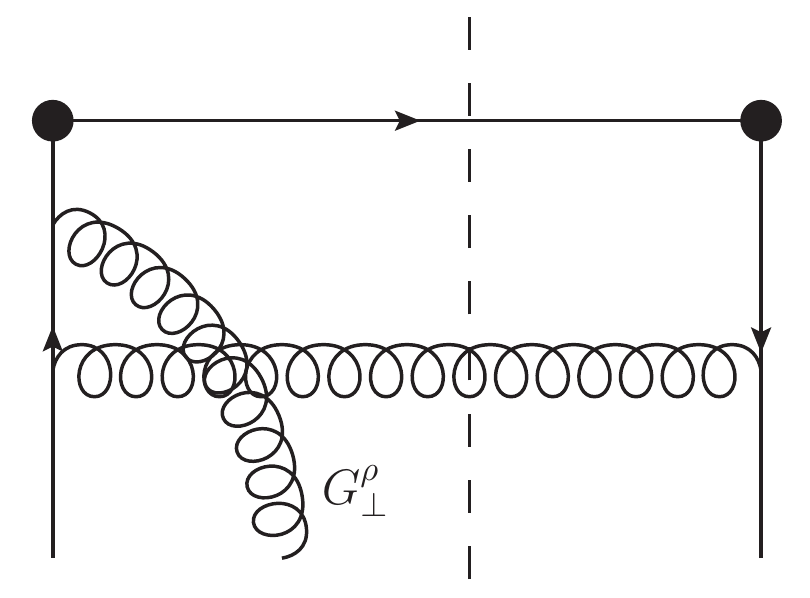}\\
(a)
\end{minipage}
\begin{minipage}[b]{0.2\textwidth}
\includegraphics[width=\textwidth]{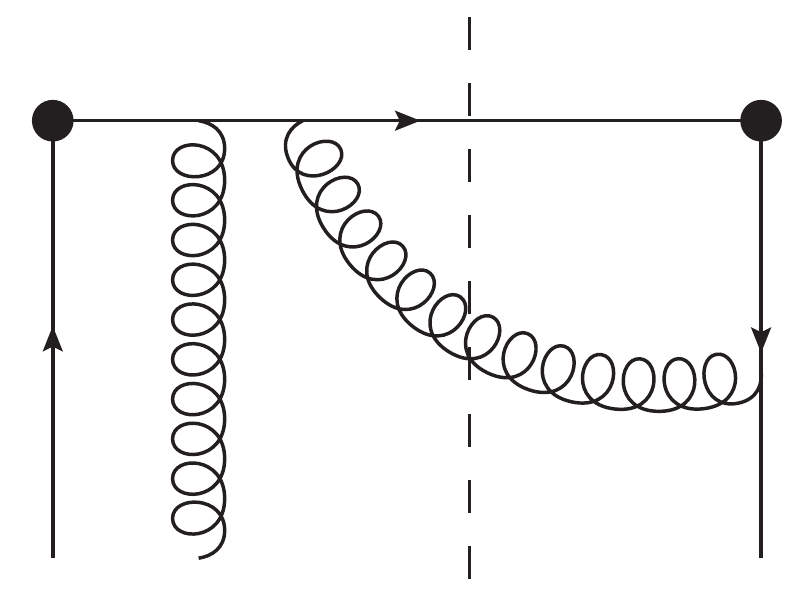}\\
(b)
\end{minipage}
\begin{minipage}[b]{0.2\textwidth}
\includegraphics[width=\textwidth]{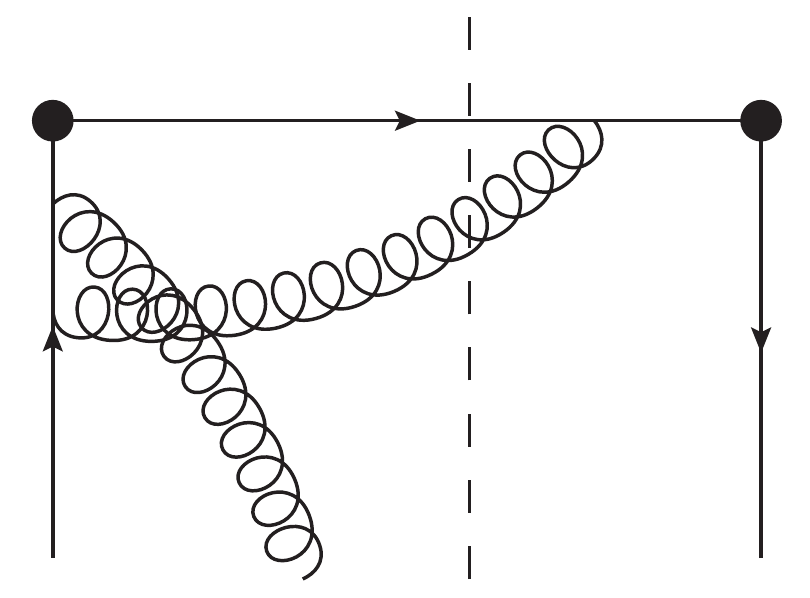}\\
(c)
\end{minipage}
\begin{minipage}[b]{0.2\textwidth}
\includegraphics[width=\textwidth]{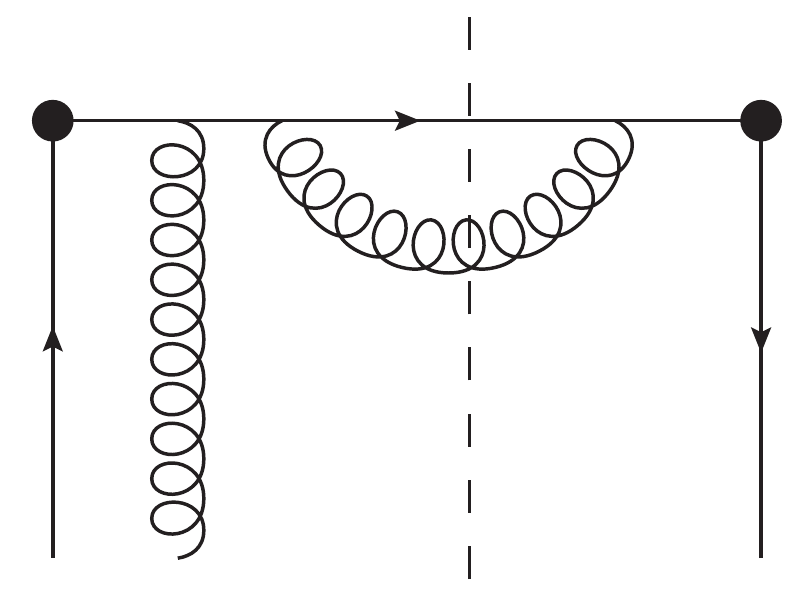}\\
(d)
\end{minipage}
\end{center}
\caption{Diagrams contributing to SGP with transverse gluon insertion. }
\label{fig:Gperp}
\end{figure}

Next, we consider the SGP from final gluon, Fig.\ref{fig:SSA}(b). The SGP appears
in the propagator $-i/[(k_g-k)^2+i\ep]$.
Explicitly, Fig.\ref{fig:SSA}(b) is proportional to
\begin{align}
-gf^{abc}\Gamma_{\al\be\ga}(k,-k_g,k_g-k)G_a^\al\frac{-i}{(k_g-k)^2+i\ep}.
\end{align}
The three-gluon vertex can be written into two parts:
\begin{align}
\Gamma_{\al\be\ga}(k,-k_g,k_g-k)=&\Big(g_{\al\be}(k_g-k)_\ga
+g_{\be\ga}k_\al+g_{\ga\al}k_{g\be}\Big)
+2\Big(
g_{\al\be}k_\ga-g_{\be\ga}k_{g\al}-g_{\ga\al}k_\be\Big).
\end{align}
The first part is identified as scalar gluon contribution, which cancels between
different diagrams due to Ward identity $\langle M|\partial\cdot G|N\rangle=0$ for
physical states $|M\rangle, |N\rangle$.
For example, $(k_g-k)_\ga$ term of Fig.\ref{fig:SSA}(b) will be cancelled
by the same term of Fig.\ref{fig:SSA}(d).
In addition, when $k_\al$ is contracted with $G_a^\al$,
it produces a contribution of order $\la^2$, i.e., $k_\al G_a^\al=O(\la^2)$. When
$k_{g\be}$ contracts with $P^{\be\be'}(k_g)$, it just vanishes
because $k_{g\be}P^{\be\be'}=0$. Now, it is clear that the scalar part in three-gluon vertex can be ignored. Then,
\begin{align}
-gf^{abc}P^{\be\be'}(k_g)\Gamma_{\al\be\ga}(k,-k_g,k_g-k)G_a^\al
\frac{-i}{(k_g-k)^2+i\ep}
\doteq &igf^{abc}P^{\be\be'}(k_g)
\Big[
\frac{G_{a\be}k_\ga-G_{a\ga}k_\be}{-k_g^- k^+ +i\ep}
-g_{\be\ga}\frac{k_g\cdot G_a}{-k_g\cdot k+i\ep}
\Big].
\end{align}
Now $k_g\cdot G_a$ contains longitudinal and transverse gluon contributions, because $k_{g\perp}=q_\perp+k_\perp$ is an $O(1)$ quantity and $k_{g\perp}\cdot G_{a\perp}$
is still $O(\la)$. For the same reason, the
denominator should be expanded as well. The result is very interesting,
\begin{align}
\frac{k_g\cdot G_a}{-k_g\cdot k+i\ep}
=& \frac{k_g^- G_a^+}{-k_g^- k^+ +i\ep}
-k_g^- k_{g\perp}^\rho\frac{k^+ G_{a\perp\rho}-k_{\perp\rho}G_a^+}{
(-k_g^- k^+ +i\ep)^2}+O(\la^2).
\end{align}
The transverse component of gluon, $G_{a\perp}$, becomes a part of gluon field strength tensor automatically.
The associated pole is a double pole. Apparently, the first term should be
a part of gauge link. Due to the commutation relation of color matrix
$if^{abc}T^c=[T^a,T^b]$,
we have
\begin{align}
\begin{matrix}\includegraphics[width=0.2\textwidth]{ssa_b}\end{matrix}
=\frac{G_a^+k_\perp^\rho}{-k^++i\ep}
\frac{\partial}{\partial k_\perp^\rho}
\left(
\begin{matrix}
\includegraphics[width=0.2\textwidth]{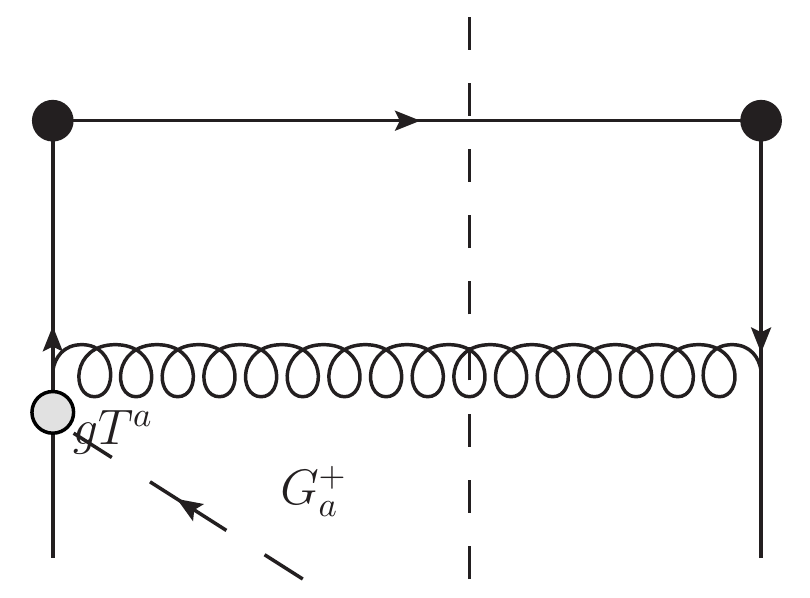}
\end{matrix}
-
\begin{matrix}
\includegraphics[width=0.2\textwidth]{k_perp_b}
\end{matrix}
\right)
+\text{field strength term}.
\end{align}
As expected, the second term in the bracket of above equation cancels the second term on RHS of eq.(\ref{eq:k_expansion}). Now, for SGP contribution with
quark fragmentation we have
\begin{align}
&\begin{matrix}\includegraphics[width=0.2\textwidth]{ssa_a}\end{matrix}
+\begin{matrix}\includegraphics[width=0.2\textwidth]{ssa_b}\end{matrix}
\no
=& \frac{G_a^+(\xi^-)k_\perp^\rho}{-k^+ +i\ep}
\begin{matrix}
\includegraphics[width=0.2\textwidth]{k_perp_c}
\end{matrix}
+\frac{G_a^+ k_\perp^\rho}{-k^++i\ep}\frac{\partial}{\partial k_\perp^\rho}
\begin{matrix}\includegraphics[width=0.2\textwidth]{k_perp_a}\end{matrix}
+\text{field strength tensor term}.\no
=&
k_\perp^\rho\frac{\partial}{\partial k_\perp^\rho}
\begin{matrix}
{\includegraphics[width=0.2\textwidth]{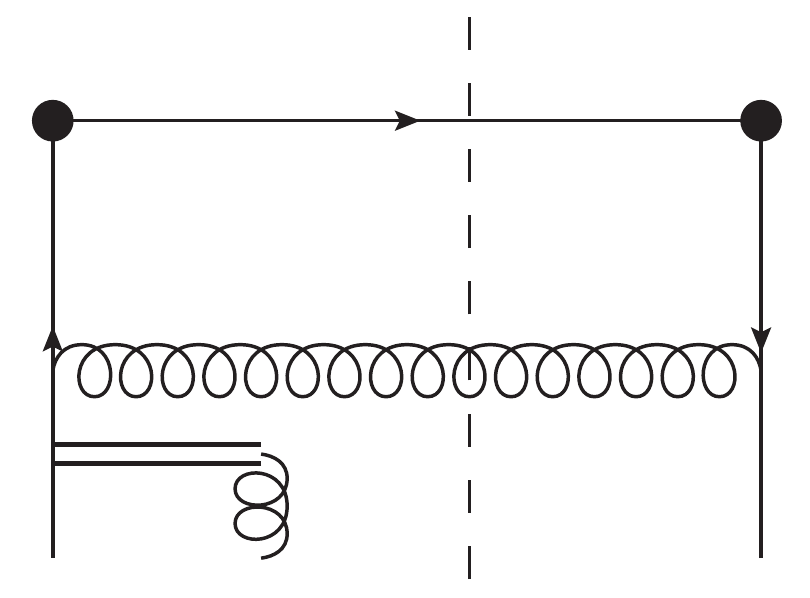}}
\end{matrix}
+\frac{G_a^+ k_\perp^\rho}{-k^++i\ep}\frac{\partial}{\partial k_\perp^\rho}
\begin{matrix}\includegraphics[width=0.2\textwidth]{k_perp_a}\end{matrix}
+\text{field strength tensor term}.
\end{align}
Now, only the second term is not in a gauge invariant form, which indicates that
our calculation for SGP is not complete. Actually, the transverse gluon also
contributes by coupling to the intermediate fermion, as shown in Fig.\ref{fig:Gperp}(a).
The SGP appears by writing
$G_\perp^\rho$ into gluon field strength tensor, i.e.,
\begin{align}
G_{a\perp}^\rho=\frac{1}{-k^++i\ep}(-k^+G_{a\perp}^\rho).
\label{eq:replacement}
\end{align}
The quantity in the bracket should be viewed as a part of gluon field strength
tensor, and the SGP is given by $1/(-k^+ +i\ep)$. Taking Fig.\ref{fig:Gperp}(a)
into account, we have
\begin{align}
\frac{G_a^+ k_\perp^\rho}{-k^+ +i\ep}
\frac{\partial}{\partial k_\perp^\rho}
\begin{matrix}
\includegraphics[width=0.2\textwidth]{k_perp_a}
\end{matrix}
+
\begin{matrix}\includegraphics[width=0.2\textwidth]{G_perp_a}\end{matrix}
=
\frac{-k^+G_{a\perp}^\rho+k_\perp^\rho G_a^+}{-k^+ +i\ep}
\begin{matrix}
{\includegraphics[width=0.2\textwidth]{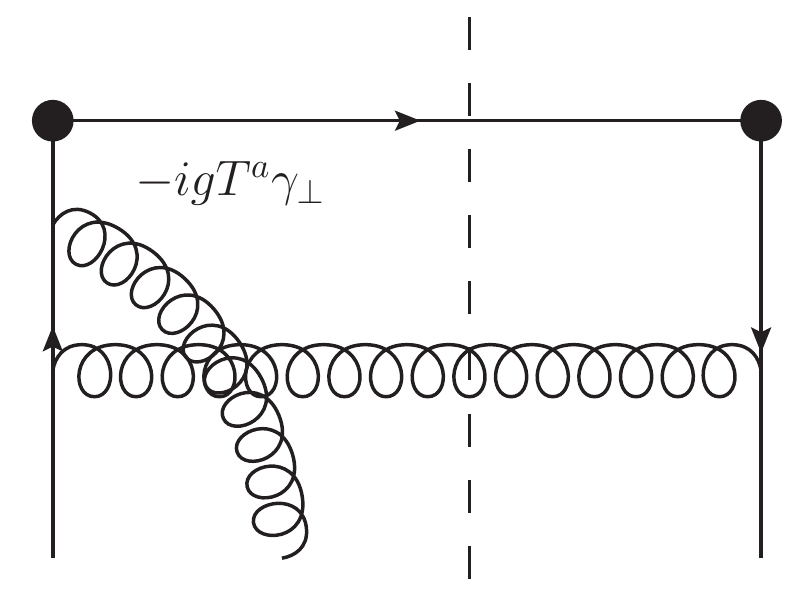}}
\end{matrix}
\label{eq:derivative_transverse_gluon}
\end{align}
In this way, gluon field strength tensor is retained by adding transverse gluon
contribution.
Finally, we have
\begin{align}
&\begin{matrix}\includegraphics[width=0.2\textwidth]{ssa_a}\end{matrix}
+\begin{matrix}\includegraphics[width=0.2\textwidth]{ssa_b}\end{matrix}
+\begin{matrix}\includegraphics[width=0.2\textwidth]{G_perp_a}\end{matrix}\no
=&
k_\perp^\rho\frac{\partial}{\partial k_\perp^\rho}
\begin{matrix}
{\includegraphics[width=0.2\textwidth]{tensor_a}}
\end{matrix}
+\text{field strength tensor term}.
\label{eq:diagram_sum}
\end{align}
In the above, the first term with transverse derivative acting on gauge link can
be written as
\begin{align}
w^{\mu\nu}\supset &
\int dk_1^+\int dk^+\frac{1}{-k^+ +i\ep}
\frac{\partial H^{\mu\nu}(k_1+k)}{\partial k_\perp^\rho}
\int\frac{d\xi^-d\xi_1^-}{(2\pi)^2}e^{ik^+\xi^-+ik_1^+\xi_1^-}
g_s\langle Ps_\perp|
\bar{\psi}(0)\ga^+ i[\partial_\perp^\rho G^+(\xi^-)]
\psi(\xi_1^-)|Ps_\perp\rangle,\no
H^{\mu\nu}=& \frac{1}{4N_c}
\int \frac{d^n k_g}{(2\pi)^n}(2\pi)
\de(k_g^2)\de^n(k+k_1+q-p_h-k_g)P^{\be\be'}(k_g)\no
&\times Tr\Big[\ga^-(igT^b\ga_{\be'})\frac{-i}{\s{p}_h-\s{q}-i\ep}
\ga^\nu\s{p}_h\ga^\mu\frac{i}{\s{p}_h-\s{q}+i\ep}(-igT^b\ga_\be)\Big].
\label{eq:reduced_hard}
\end{align}
The factor $1/(4N_c)$ comes from the projection of quark-gluon-quark correlation
function, and the trace in $H^{\mu\nu}$ includes color trace.
Since $H^{\mu\nu}$ depends on $k_1+k$ only, one can make a variable transformation
$k_1^+\rightarrow k_1^+-k^+$ to eliminate $k^+$ in $H^{\mu\nu}$. Then integrating over $k^+$ gives
\begin{align}
w^{\mu\nu}\supset &
(-2\pi i)\int dk_1^+
\frac{\partial H^{\mu\nu}(k_1)}{\partial k_{1\perp}^\rho}
\int\frac{d\xi_1^-}{(2\pi)^2}e^{ik_1^+\xi_1^-}
\langle Ps_\perp|
\bar{\psi}(0)\ga^+ \Big[\partial_\perp^\rho ig_s\int^\infty_{\xi_1^-}d\xi^- G^+(\xi^-)\Big]\psi(\xi_1^-)|Ps_\perp\rangle\no
=&
i\int dk_1^+
\frac{\partial H^{\mu\nu}(k_1)}{\partial k_{1\perp}^\rho}
\int\frac{d\xi_1^-}{2\pi}e^{ik_1^+\xi_1^-}
\langle Ps_\perp|
\bar{\psi}(0)\ga^+ \Big[\partial_\perp^\rho \mathcal{L}_n(\xi_1^-)
\Big]\psi(\xi_1^-)|Ps_\perp\rangle,
\end{align}
where the gauge link $\mathcal{L}_n$ is defined in eq.(\ref{eq:gauge_link}).

Next we consider the expansion in $k_{1\perp}$. Since $k_{1\perp}\sim O(\la)$,
only $G^+$ contributes. Because $k_\perp=0$, it is clear that such $G^+$ in Fig.\ref{fig:SSA}(a,b) are combined into a gauge link.
Then, $k_{1\perp}$ expansion gives
\begin{align}
w^{\mu\nu}\supset &
\int dk_1^+\int dk^+\frac{1}{-k^+ +i\ep}
\frac{\partial H^{\mu\nu}(k_1+k)}{\partial k_{1\perp}^\rho}
\int\frac{d\xi^-d\xi_1^-}{(2\pi)^2}e^{ik^+\xi^-+ik_1^+\xi_1^-}
g_s\langle Ps_\perp|
\bar{\psi}(0)\ga^+ G^+(\xi^-)i\partial_\perp^\rho \psi(\xi_1^-)|Ps_\perp\rangle\no
=& i\int dk_1^+
\frac{\partial H^{\mu\nu}(k_1)}{\partial k_{1\perp}^\rho}
\int\frac{d\xi_1^-}{2\pi}e^{ik_1^+\xi_1^-}
\langle Ps_\perp|
\bar{\psi}(0)\ga^+ \mathcal{L}_n(\xi_1^-)
\Big[ \partial_\perp^\rho\psi(\xi_1^-)\Big]|Ps_\perp\rangle,
\label{eq:reduced_hard_2}
\end{align}
with the same $H^{\mu\nu}$ in eq.(\ref{eq:reduced_hard}).
Then,
eq.(\ref{eq:reduced_hard}) and eq.(\ref{eq:reduced_hard_2})
can be combined together to give a gauge invariant
matrix element, i.e.,
\begin{align}
w^{\mu\nu}\supset &
-\int dk_1^+
\frac{\partial H^{\mu\nu}(k_1)}{\partial k_{1\perp}^\rho}
\int\frac{d\xi_1^-}{2\pi}e^{ik_1^+\xi_1^-}
\langle Ps_\perp|
\bar{\psi}(0)\ga^+ i\partial_\perp^\rho\Big[\mathcal{L}_n(\xi_1^-)
\psi(\xi_1^-)\Big]|Ps_\perp\rangle.
\end{align}
The correlation function appearing on RHS above is rightly $q_\partial(x_1)$,
which is gauge invariant.

On the other hand, the field strength tensor term in eq.(\ref{eq:diagram_sum}) is proportional to $T_F(x_1,x_1+x)$. 
But since Fig.\ref{fig:SSA} contains a double pole, a derivative in $x$ acting on $T_F(x_1,x_1+x)$ will appear. The double pole contribution is
\begin{align}
\int dx T_F(x_1,x_1+x)H(x)\frac{1}{(x-i\ep)^2},
\end{align}
where $H(x)$ is the hard coefficient. Other variables are suppressed for simplicity.
By using integration by part, it becomes
\begin{align}
&\int dx\frac{d [T_F(x_1,x_1+x)H(x)]}{dx}\frac{1}{x-i\ep}
\doteq i\pi \frac{d}{dx}\Big[T_F(x_1,x_1+x)H(x)\Big]_{x=0}\no
=&i\pi \Big[\frac{1}{2}H(x=0)\frac{d}{d x_1}T_F(x_1,x_1)
+T_F(x_1,x_1)\frac{dH(x=0)}{dx}\Big].
\end{align}
Thus $d T_F(x_1,x_1)/d x_1$ appears.
In the above we have used the symmetry $T_F(x_1,x_2)=T_F(x_2,x_1)$, and the principal integration is ignored because it does not contribute to the real
part of $w^{\mu\nu}$.

The same derivation can be applied to Fig.\ref{fig:SSA}(e,f) directly. The
sum of Fig.\ref{fig:Gperp}(c) and Fig.\ref{fig:SSA}(e,f) can be decomposed
into gauge link contribution and gluon field strength tensor contribution,
similar to eq.(\ref{eq:diagram_sum}). For Fig.\ref{fig:SSA}(c,d), the
$k_\perp$ expansion gives
\begin{align}
&\begin{matrix}\includegraphics[width=0.15\textwidth]{ssa_c}\end{matrix}
+\begin{matrix}\includegraphics[width=0.15\textwidth]{ssa_d}\end{matrix}
=\frac{G_a^+(\xi^-)k_\perp^\rho}{-k^+ +i\ep}
\frac{\partial}{\partial k_\perp^\rho}
\begin{matrix}
\includegraphics[width=0.15\textwidth]{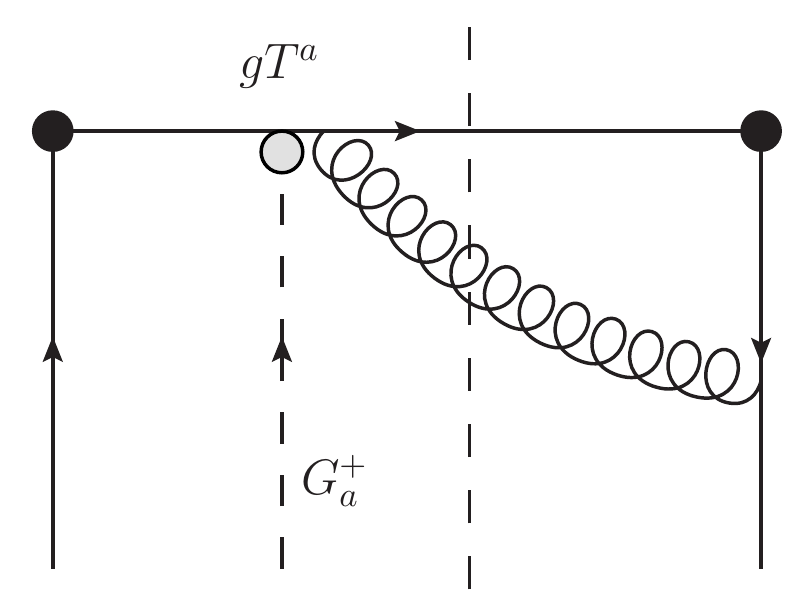}
\end{matrix}
+\text{field strength tensor term}.\no
&\begin{matrix}\includegraphics[width=0.15\textwidth]{scalar_cd_1}\end{matrix}
=
\begin{matrix}\includegraphics[width=0.15\textwidth]{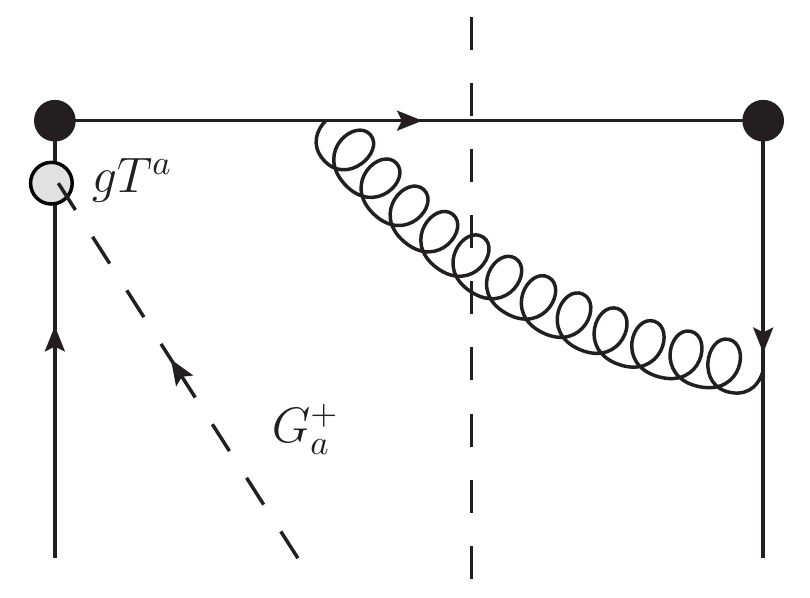}\end{matrix}
-
\begin{matrix}\includegraphics[width=0.15\textwidth]{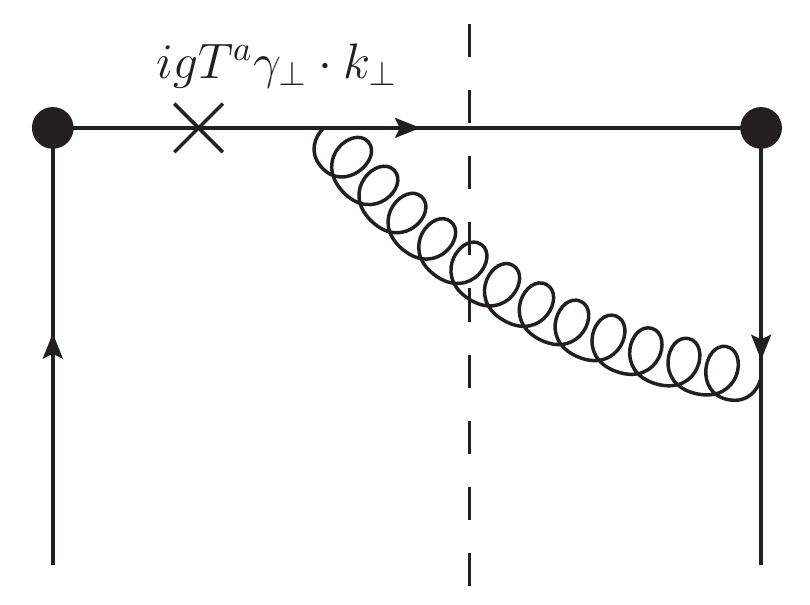}\end{matrix},
\end{align}
Notice that there is a minus sign before the second diagram on RHS of the second
line. This minus sign and the vertex $igT^a\s{k}_\perp$ together give the standard
quark-gluon vertex. In this way, this diagram is combined with Fig.\ref{fig:Gperp}(c) to give a field strength tensor. The final result is
\begin{align}
&\begin{matrix}\includegraphics[width=0.15\textwidth]{ssa_c}\end{matrix}
+\begin{matrix}\includegraphics[width=0.15\textwidth]{ssa_d}\end{matrix}
+\begin{matrix}\includegraphics[width=0.15\textwidth]{G_perp_b}\end{matrix}
=k_\perp^\rho
\frac{\partial}{\partial k_\perp^\rho}
\begin{matrix}
\includegraphics[width=0.15\textwidth]{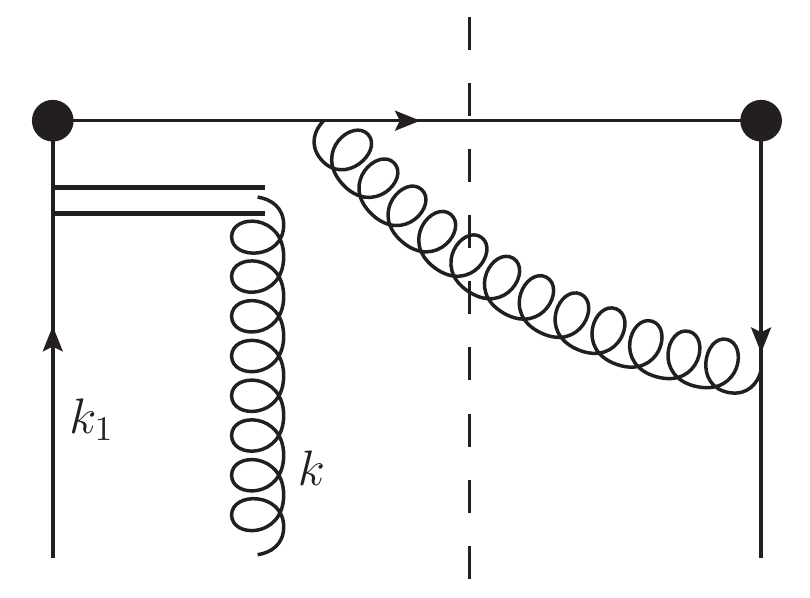}
\end{matrix}
+\text{field strength tensor term}.
\end{align}
The expansion in
$k_{1\perp}$ is the same as Fig.\ref{fig:SSA}(a,b). As a result, the contribution
from Fig.\ref{fig:SSA}(c,d) and Fig.\ref{fig:Gperp}(b) can be expressed by
$q_\partial$ and $T_F$.
Fig.\ref{fig:SSA}(g,h) and Fig.\ref{fig:Gperp}(d) can be treated in the same way.

\begin{figure}
\begin{center}
\begin{minipage}[b]{0.2\textwidth}
\includegraphics[width=\textwidth]{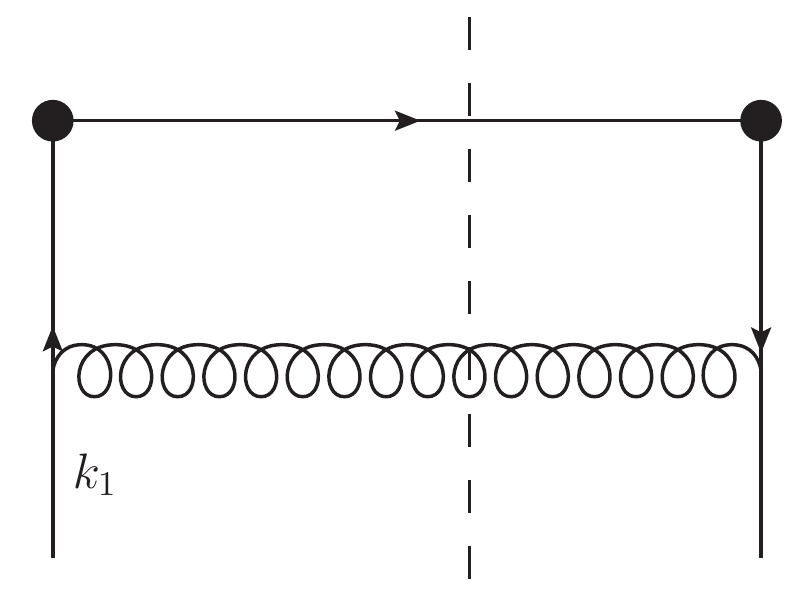}\\
(a)
\end{minipage}
\begin{minipage}[b]{0.2\textwidth}
\includegraphics[width=\textwidth]{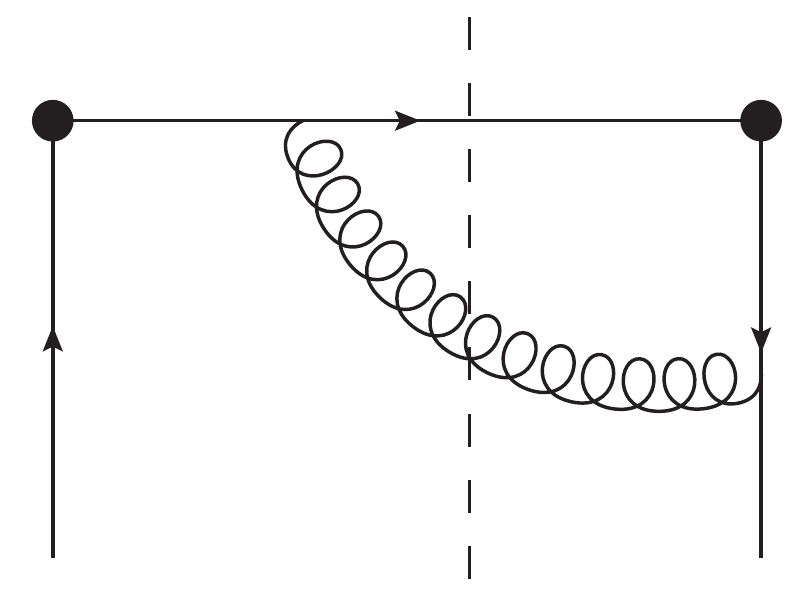}\\
(b)
\end{minipage}
\begin{minipage}[b]{0.2\textwidth}
\includegraphics[width=\textwidth]{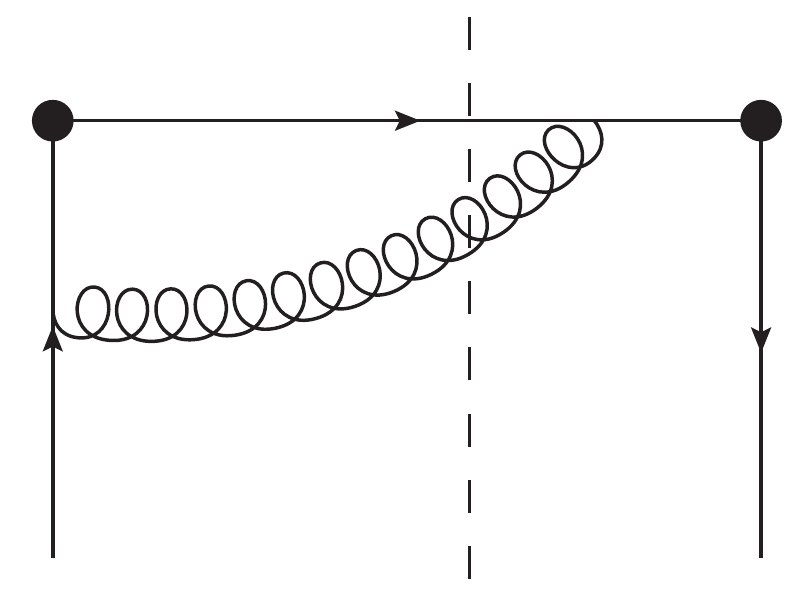}\\
(c)
\end{minipage}
\begin{minipage}[b]{0.2\textwidth}
\includegraphics[width=\textwidth]{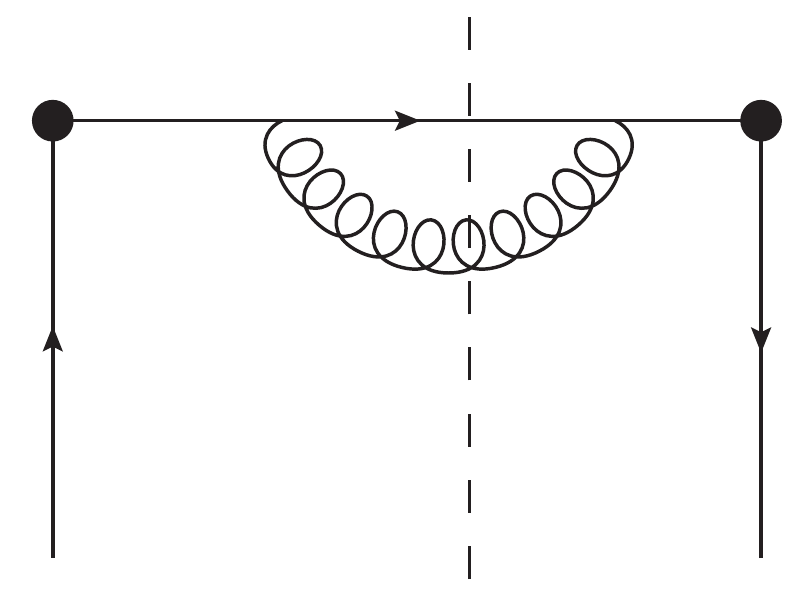}\\
(d)
\end{minipage}
\end{center}
\caption{Diagrams which can give the coefficient $H_{ij}^{\mu\nu}$ in eq.(\ref{eq:H_12}).}
\label{fig:no_gluon}
\end{figure}

Now, we have shown that the twist expansion can generate $q_\partial$ and $T_F$
in a very natural way. Especially, $T_F$ part can be recovered by transverse gluon $G_\perp$ definitely. Our formula for SGP contribution with quark fragmenting
thus is
\begin{align}
w^{\mu\nu}=&\int dx_1\Big\{
\frac{\partial H_1^{\mu\nu}(k_{1\perp})}{\partial {k_{1\perp}^\rho}}
\tilde{s}_\perp^\rho q_\partial(x_1)
-\frac{i}{\pi}
\int dx \frac{H_{2\perp\rho}^{\mu\nu}(x_1,x)}{(x-i\ep)^2}\tilde{s}_\perp^\rho
T_F(x_1,x_1+x)\Big\},
\label{eq:main}
\end{align}
where $H_{1,2}$ are particular hard coefficients defined as follows:
\begin{align}
H_1^{\mu\nu}=& H^{\mu\nu}_{ij}[\frac{1}{2N_c}\ga^- p_A^+]_{ji},\
H_{2\perp\rho}^{\mu\nu}=-\pi x [H_{a\perp\rho}^{\mu\nu}]_{ij}
[\frac{\ga^- p_A^+ T^a}{4\pi N_c C_F}]_{ji},
\label{eq:H_12}
\end{align}
where $H^{\mu\nu}_{ij}$ is the hard part of Fig.\ref{fig:no_gluon}, which contains no initial gluon, and $[H_{a\perp\rho}^{\mu\nu}]_{ij}$ is the hard part of
Fig.\ref{fig:Gperp}, in which the initial gluon is a transverse one.
This is our main formula.
From this formula, it is clear that if we use
$G_\perp$ only to derive factorization formula, $q_\partial$ contribution
is missing. And if we use $q_\partial$ only, $T_F$ contribution
is missing. But if we use $G^+$ to derive, both contributions can be included.
This is the reason why $G^+$ can give correct hard coefficients.

In addition, the same derivation can be applied to the case where gluon is the
fragmenting parton, where the double poles are given by Fig.\ref{fig:SSA}(a,c,e,g).
The hard coefficients can be obtained in a transparent way, by changing the
projection operators to those for gluon fragmentation functions. Eq.(\ref{eq:H_12})
is still right for gluon as fragmentation parton.

\section{Hard coefficients}
In our main formula eq.(\ref{eq:main}) for quark fragmenting, the delta function $\de(k_g^2)$ for on-shell final gluon is contained in $H_{1}^{\mu\nu}$ and $H_{2,a\perp\rho}^{\mu\nu}$. It is helpful to write out the delta function explicitly in practical calculation, since the derivative also acts on the delta function. In addition, the double pole at $x=0$ can be eliminated by integration
by part. That is,
\begin{align}
w^{\mu\nu}=&\int dx_1 \Big\{
\frac{\partial[\tilde{H}_1^{\mu\nu}\de(k_g^2)]}{\partial k_{1\perp}^\rho}
\tilde{s}_\perp^\rho q_\partial(x_1)
-\frac{i}{\pi}\int dx \frac{\tilde{H}_{2\perp\rho}^{\mu\nu}(x_1,x)\de(k_g^2)}{
(x-i\ep)^2}\tilde{s}_\perp^\rho T_F(x_1,x_1+x)\Big\}\no
=& \int dx_1 \tilde{s}_\perp^\rho \Big\{
\frac{\partial[\tilde{H}_1^{\mu\nu}\de(k_g^2)]}{\partial k_{1\perp}^\rho}
q_\partial(x_1)
-\frac{i}{\pi}\int dx \frac{1}{x-i\ep}
\frac{\partial}{\partial x}\Big[\tilde{H}_{2\perp\rho}^{\mu\nu}(x_1,x)\de(k_g^2)
T_F(x_1,x_1+x)\Big]\Big\}.
\end{align}
In the first term, $k_g=x_1 p_A+q-p_h+k_{1\perp}$; in the second term,
$k_g=(x_1+x)p_A+q-p_h$. Since there is a derivative in $x$ in the second
term, it is better to make a variable transformation $x_1\rightarrow x_1-x$
so that there is no $x$ contained in $\de(k_g^2)$ any more. This trick
simplifies the calculation greatly. Then,
\begin{align}
w^{\mu\nu}
=& \int dx_1 \tilde{s}_\perp^\rho \Big\{
\frac{\partial[\tilde{H}_1^{\mu\nu}\de(\Delta+2k_{1\perp}\cdot q_\perp)]}{
\partial k_{1\perp}^\rho}
q_\partial(x_1)
-\frac{i}{\pi}\int dx \frac{1}{x-i\ep}
\frac{\partial}{\partial x}\Big[\tilde{H}_{2\perp\rho}^{\mu\nu}(x_1-x,x)
\de(\Delta)T_F(x_1-x,x_1)\Big]\Big\},
\end{align}
with
\begin{align}
\Delta=(x_1 p_A+q-p_h)^2=\hat{z}\Big[\frac{(1-\hat{z})(1-\hat{x}_1)}{\hat{z}\hat{x}_1}-q_T^2\Big].
\end{align}
Taking the pole contribution in $x$ integration, we have
\begin{align}
w^{\mu\nu}
=& \int dx_1 \tilde{s}_\perp^\rho \Big\{
\frac{\partial[\tilde{H}_1^{\mu\nu}\de(\Delta+2k_{1\perp}\cdot q_\perp)]}{
\partial k_{1\perp}^\rho}
q_\partial(x_1)
+\frac{\partial}{\partial x}\Big[\tilde{H}_{2\perp\rho}^{\mu\nu}(x_1-x,x)
\de(\Delta)T_F(x_1-x,x_1)\Big]_{x=0}\Big\}.
\end{align}
In the first term the derivative also acts on the delta function.
Following equation is helpful:
\begin{align}
\frac{\partial\de(\Delta+2k_{1\perp}\cdot q_\perp)}{\partial k_{1\perp}^\rho}
=& 2q_\perp^\rho\de'(\Delta)
=\frac{q_\perp^\rho}{p_A\cdot(q-p_h)}\frac{\partial\de(\Delta)}{
\partial x_1}.
\end{align}
After integration by part, the final formula is
\begin{align}
w^{\mu\nu}
=& \int dx_1 \de(\Delta)\tilde{s}_\perp^\rho \Big\{
\Big[\frac{\partial\tilde{H}_1^{\mu\nu}}{
\partial k_{1\perp}^\rho}
-\frac{\partial}{\partial x_1}\Big(
\frac{q_\perp^\rho \tilde{H}_1^{\mu\nu}}{p_A\cdot(q-p_h)}\Big)
\Big]q_\partial(x_1)
-\frac{q_\perp^\rho \tilde{H}_1^{\mu\nu}}{p_A\cdot(q-p_h)}
\frac{\partial q_\partial(x_1)}{\partial x_1}\no
&+\Big[
T_F(x_1,x_1)\Big(\frac{\partial \tilde{H}_{2\perp\rho}^{\mu\nu}(x_1-x,x)}{\partial x}
\Big)_{x=0}
-\frac{1}{2}\tilde{H}_{2\perp\rho}^{\mu\nu}(x_1,0)
\frac{\partial}{\partial x_1}T_F(x_1,x_1)\Big]\Big\}.
\end{align}
Then, projected hadronic tensors $B_k$ defined in eq.(\ref{eq:project_W}) are
\begin{align}
B_k=&
\int \frac{dz}{z^2}D(z)\int dx_1\de(\Delta)q_\perp\cdot \tilde{s}_\perp\Big\{
C_1^k q_\partial(x_1)+C_2^k \frac{\partial}{\partial x_1}q_\partial(x_1)
+C_3^k T_F(x_1,x_1)+C_4^k\frac{\partial}{\partial x_1}T_F(x_1,x_1)
\Big\},
\end{align}
with $k=1,\cdots,4$ and
\begin{align}
C_1^k=&-\frac{q_\perp^\rho}{q_T^2}\Big[
\frac{\partial\tilde{H}_1^{k}}{
\partial k_{1\perp}^\rho}
-\frac{\partial}{\partial x_1}\Big(
\frac{q_{\perp\rho} \tilde{H}_1^{k}}{p_A\cdot(q-p_h)}\Big)
\Big],\no
C_2^k=& -\frac{\tilde{H}_1^k}{p_A\cdot(q-p_h)},\no
C_3^k=&-\frac{1}{q_T^2}q_\perp^\rho
\Big(\frac{\partial \tilde{H}_{2\perp\rho}^{k}(x_1-x,x)}{\partial x}
\Big)_{x=0},\no
C_4^k=&\frac{1}{q_T^2}\frac{1}{2}q_\perp^\rho
\tilde{H}_{2\perp\rho}^{k}(x_1,0).
\end{align}
There is a relation between $q_\partial(x)$ and $T_F(x,x)$, i.e.,
$-2q_\partial(x)=T_F(x,x)$,
which was found by many authors in different ways\cite{Boer:2003cm,Ma:2003ut,Bacchetta:2005xk,Ma:2014uma}.
With this relation, $B_k$ can be expressed by $T_F$ solely. That is,
\begin{align}
B_k=&
\int \frac{dz}{z^2}D(z)\int dx_1\de(\Delta)q_\perp\cdot \tilde{s}_\perp\Big\{
E_1^k T_F(x_1,x_1)+E_2^k\frac{\partial}{\partial x_1}T_F(x_1,x_1)
\Big\},
\end{align}
with
\begin{align}
E_1^k=-\frac{1}{2}C_1^k+C_3^k,\ E_2^k=-\frac{1}{2}C_2^k+C_4^k.
\end{align}
Our results are
\begin{align}
\Big[\frac{\al_s}{2\pi^2 N_c Q^2}\frac{z}{z_f x_1}\Big]^{-1}E_1^1=&
-\frac{{\hat{x}_1} {\hat{z}} \left(2 {\hat{x}_1}^3 \left(6 {\hat{z}}^2-6 {\hat{z}}+1\right)+{\hat{x}_1}^2 \left(-24
   {\hat{z}}^2+26 {\hat{z}}-5\right)+4 {\hat{x}_1} \left(3 {\hat{z}}^2-4 {\hat{z}}+1\right)-{\hat{z}}^2+2
   {\hat{z}}-2\right)}{({\hat{x}_1}-1)^2 ({\hat{z}}-1)^2},\no
\Big[\frac{\al_s}{2\pi^2 N_c Q^2}\frac{z}{z_f x_1}\Big]^{-1}E_1^2=&
-\frac{8 {\hat{x}_1}^2 {\hat{z}}^2}{{\hat{z}}-1},\no
\Big[\frac{\al_s}{2\pi^2 N_c Q^2}\frac{z}{z_f x_1}\Big]^{-1}E_1^3=&
-\frac{Q}{q_T}
\frac{{\hat{x}_1} {\hat{z}} \left({\hat{x}_1}^2 (8 {\hat{z}}-4)+{\hat{x}_1} (7-12 {\hat{z}})+3
   ({\hat{z}}-1)\right)}{({\hat{x}_1}-1) ({\hat{z}}-1)},\no
\Big[\frac{\al_s}{2\pi^2 N_c Q^2}\frac{z}{z_f x_1}\Big]^{-1}E_1^4=&
-\frac{4 {\hat{x}_1}^2 {\hat{z}}^2}{{\hat{z}}-1};
\end{align}
And
\begin{align}
\Big[\frac{\al_s}{2\pi^2 N_c Q^2}\frac{z {x_B}}{z_f x_1}\Big]^{-1}E_2^1=&
\frac{ {\hat{z}} \left({\hat{x}_1}^2 \left(6 {\hat{z}}^2-6 {\hat{z}}+1\right)+{\hat{x}_1} \left(-6
   {\hat{z}}^2+8 {\hat{z}}-2\right)+{\hat{z}}^2-2 {\hat{z}}+2\right)}{({\hat{x}_1}-1) ({\hat{z}}-1)^2},\no
\Big[\frac{\al_s}{2\pi^2 N_c Q^2}\frac{z{x_B}}{z_f x_1}\Big]^{-1}E_2^2=&
\frac{4 {\hat{x}_1} {\hat{z}}^2}{{\hat{z}}-1},\no
\Big[\frac{\al_s}{2\pi^2 N_c Q^2}\frac{z{x_B}}{z_f x_1}\Big]^{-1}E_2^3=&
\frac{Q}{q_T}\frac{2 {\hat{z}} ({\hat{x}_1} (2 {\hat{z}}-1)-{\hat{z}}+1)}{{\hat{z}}-1},\no
\Big[\frac{\al_s}{2\pi^2 N_c Q^2}\frac{z{x_B}}{z_f x_1}\Big]^{-1}E_2^4=&
\frac{2 {\hat{x}_1} {\hat{z}}^2}{{\hat{z}}-1},
\end{align}
with
\begin{align}
q_T=Q\sqrt{\frac{(1-\hat{z})(1-\hat{x}_1)}{\hat{z}\hat{x}_1}},\
\hat{z}=\frac{z_f}{z},\ \hat{x}_1=\frac{x_B}{x_1}.
\end{align}
These results are the same as those given in \cite{Eguchi:2006mc}.

For gluon fragmentation, the formulas are the same, but now the momentum of
final gluon is related to observed hadron by $k_g=p_H/z$, and the
corresponding hard coefficients are
\begin{align}
\Big[\frac{\al_s N_c}{2\pi^2 Q^2 }\frac{z}{z_f x_1}\Big]^{-1}E_1^1=&
\frac{{\hat{x}_1} \left(-2 {\hat{x}_1}^3 \left(6 {\hat{z}}^2-6 {\hat{z}}+1\right)+{\hat{x}_1}^2 \left(24 {\hat{z}}^2-22
   {\hat{z}}+3\right)+4 {\hat{x}_1} (2-3 {\hat{z}}) {\hat{z}}+{\hat{z}}^2+1\right)}{({\hat{x}_1}-1)^2 ({\hat{z}}-1)},\no
\Big[\frac{\al_s N_c}{2\pi^2 Q^2 }\frac{z}{z_f x_1}\Big]^{-1}E_1^2=&
-8 {\hat{x}_1}^2 {\hat{z}},\no
\Big[\frac{\al_s N_c}{2\pi^2 Q^2 }\frac{z}{z_f x_1}\Big]^{-1}E_1^3=&
-\frac{Q}{q_T}
\frac{{\hat{x}_1} \left({\hat{x}_1}^2 (8 {\hat{z}}-4)+{\hat{x}_1} (5-12 {\hat{z}})+3 {\hat{z}}\right)}{{\hat{x}_1}-1},\no
\Big[\frac{\al_s N_c}{2\pi^2 Q^2 }\frac{z}{z_f x_1}\Big]^{-1}E_1^4=&
-4 {\hat{x}_1}^2 {\hat{z}};
\end{align}
and
\begin{align}
\Big[\frac{\al_s N_c}{2\pi^2 Q^2}\frac{z {x_B}}{z_f x_1}\Big]^{-1}E_2^1=&
\frac{{\hat{x}_1}^2 \left(6 {\hat{z}}^2-6 {\hat{z}}+1\right)+2 {\hat{x}_1} (2-3 {\hat{z}})
   {\hat{z}}+{\hat{z}}^2+1}{({\hat{x}_1}-1) ({\hat{z}}-1)},\no
\Big[\frac{\al_s N_c}{2\pi^2 Q^2}\frac{z {x_B}}{z_f x_1}\Big]^{-1}E_2^2=&
4 {\hat{x}_1} {\hat{z}},\no
\Big[\frac{\al_s N_c}{2\pi^2 Q^2}\frac{z {x_B}}{z_f x_1}\Big]^{-1}E_2^3=&
\frac{Q}{q_T}\Big[{\hat{x}_1} (4 {\hat{z}}-2)-2 {\hat{z}}\Big],\no
\Big[\frac{\al_s N_c}{2\pi^2 Q^2}\frac{z {x_B}}{z_f x_1}\Big]^{-1}E_2^4=&
2 {\hat{x}_1} {\hat{z}}.
\end{align}
These results are also the same as those given in \cite{Eguchi:2006mc}.

\begin{figure}
\begin{center}
\begin{minipage}[b]{0.2\textwidth}
\includegraphics[width=\textwidth]{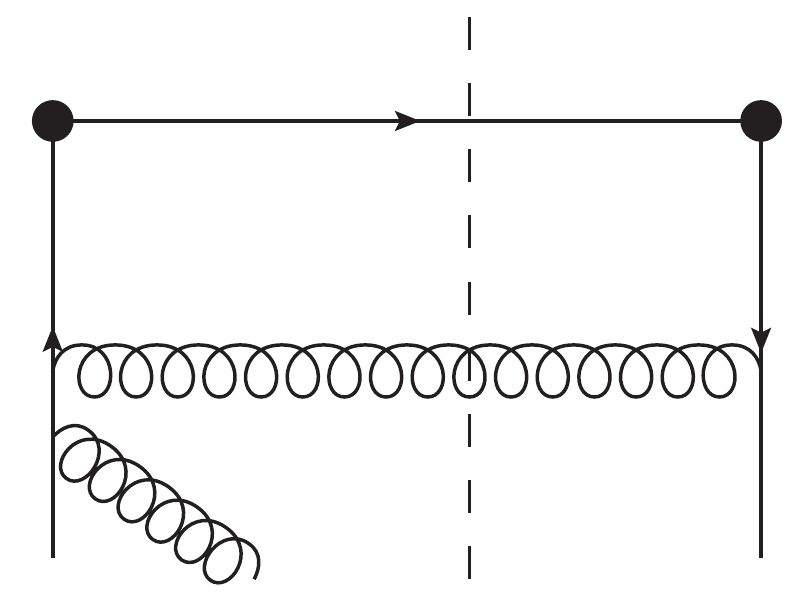}\\
(a)
\end{minipage}
\begin{minipage}[b]{0.2\textwidth}
\includegraphics[width=\textwidth]{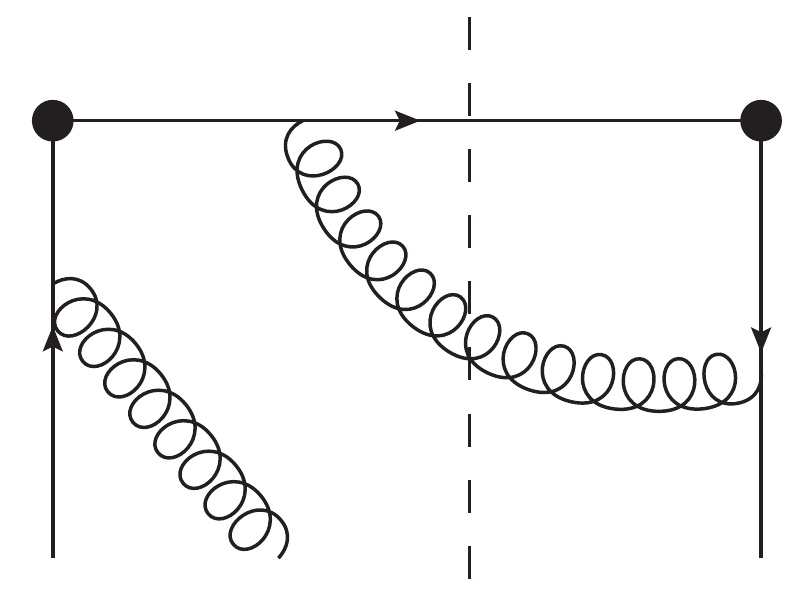}\\
(b)
\end{minipage}
\begin{minipage}[b]{0.2\textwidth}
\includegraphics[width=\textwidth]{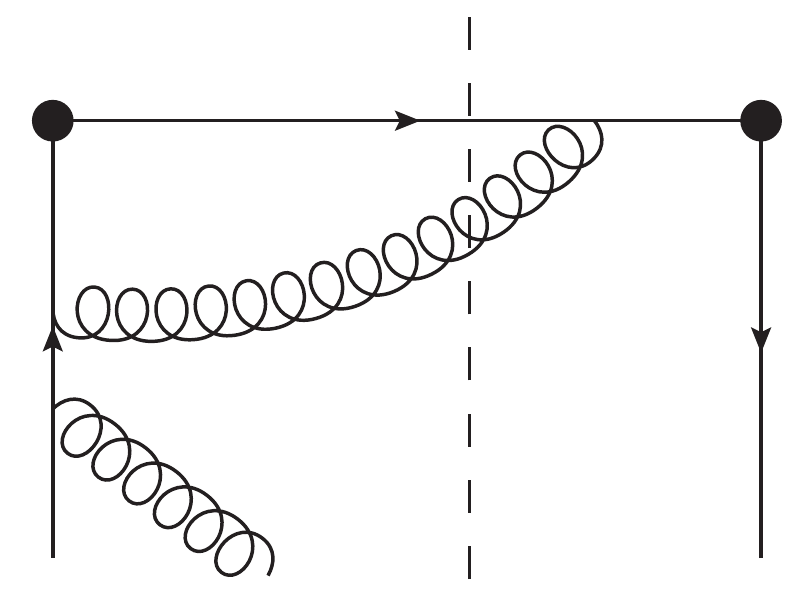}\\
(c)
\end{minipage}
\begin{minipage}[b]{0.2\textwidth}
\includegraphics[width=\textwidth]{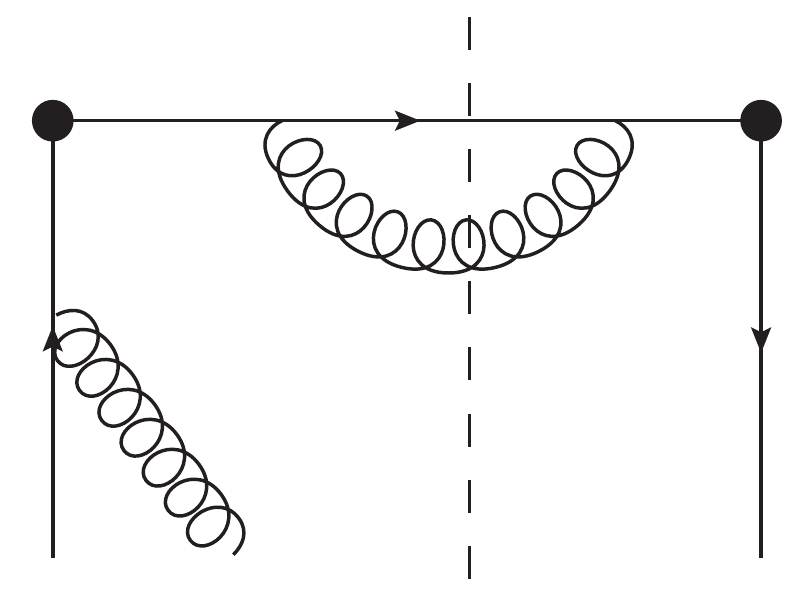}\\
(d)
\end{minipage}
\end{center}
\caption{Contact diagrams which are not included in hard coefficient $H_{2\perp\rho}^{\mu\nu}$ in our calculation.
The initial gluon is transverse in these
diagrams. Conjugated diagrams give the same result and are not shown. }
\label{fig:contact}
\end{figure}
We also do calculations in light-cone gauge with $G^+=0$ or $p_H\cdot G=0$.
For the case of gluon fragmentation, the final gluon is transversely polarized
in both Feynman and light-cone gauges. Thus, in this case
the hard coefficients in these two gauges are the same.
For the case of quark fragmentation,
$C_2^k$ and $C_4^k$ in light-cone gauge are the same as those in Feynman gauge, but $C_1^k$ and $C_3^k$ are different. The difference $\Delta C_i^k=
C_i^k|_{LC}-C_i^k|_{Feynman}$ are shown in Appendix.\ref{sec:c_i}.
Since $C_2$ is determined
by the on-shell $\ga^* q$ amplitude, $C_2$ is gauge independent naturally.
The reason that $C_3^k$
depends on gauge is $\tilde{H}_{2\perp\rho}^k$ is not given by a complete
amplitude for the scattering of $\ga^*$ and $qg$, because the contact diagrams, Fig.\ref{fig:contact}, are not included.
In collinear expansion, the contribution of these contact diagrams
has been included in the contribution of $q_\partial$, because the intermediate
propagator with momentum $k_1+k$ is collinear. Since $q_\partial$ contribution
has been taken into account, including these contact diagrams in $\tilde{H}_{2\perp\rho}^k$ is a double counting. These contact diagrams are
regular at $x=0$, and thus vanish when $x\rightarrow 0$ in $C_4^k$ because of eq.(\ref{eq:H_12}), where $x$ is multiplied to the amplitude in $\tilde{H}_{2\perp\rho}^{k}$. So, $C_4^k$ is gauge invariant, even though the
on-shell amplitude is incomplete. For $C_3^k$,
due to the derivative in $x$, contact diagrams give nonzero contribution.
Moreover, this contribution depends on gauge. The gauge dependence can be
seen from following simple analysis.
\begin{align}
\tilde{H}_{2\perp\rho}^{\mu\nu}|_{contact}\propto &
\text{Tr}\Big[p_A^+\ga^-(-igT^a)\s{G}_{a\perp}(\xi^-)\frac{i}{\s{k}_1+\s{k}+i\ep}
(\cdots)\Big]\no
=& \text{Tr}\Big[p_A^+\ga^-(-igT^a)\s{G}_{a\perp}(\xi^-)\frac{i\ga^+}{2(k_1+k)^+}
(\cdots)\Big],
\end{align}
where $(\cdots)$ represent $\ga^* q$ scattering cross section. If $(\cdots)$ is
projected by $p_A^+\ga^-$, the resulting $\ga^* q$ cross section
is on-shell and thus gauge independent.
But here, as we can see, it is projected by $p_A^+\ga^-\ga_\perp\ga^+$, which gives
a kind of off-shell contribution, thus the trace depends on gauge. Since the complete amplitude is gauge invariant, the absence of compact diagrams gives
gauge dependent $C_3^k$.
However, the gauge dependence in $C_1^k$ and $C_3^k$ cancel each other, and the
final coefficient $E_3^k$ does not depend on gauge. This also supports the
conclusion that $q_\partial(x)$ and $T_F(x,x)$ are not independent.

\section{Discussion and Summary}
Before the summary we want to compare our formalism with that given in \cite{Eguchi:2006mc}.
The main difference
is in our calculation the non-pole diagrams in Fig.\ref{fig:Gperp} for
$G_\perp$ contribution are included.
In \cite{Eguchi:2006mc}, these diagrams are ignored. Really, if
one changes $G_\perp$ in these diagrams to $G^+$, these diagrams do not give
pole contribution and should be dropped. In addition, in our calculation,
the cancellation between mirror diagrams are not used. The diagram with coherent
gluon on RHS of the cut is taken as conjugated diagrams, and from PT symmetry one
can show these conjugated diagrams give the same results, thus are not needed
to be calculated again. But in the formalism of \cite{Eguchi:2006mc}, these conjugated diagrams give different
results and have to be calculated separately.
At last, in our calculation it is shown explicitly how the gauge link is reproduced
at twist-3 level. Different from twist-2 factorization, the gauge link cannot be
obtained by using Ward identity solely: Part of $G^+$ has to be combined with
$G_\perp$ to form gluon field strength tensor.

The equivalence between transverse momentum expansion of fermion propagator and
the insertion of a transverse gluon is very general. Thus, the
derivation is possible to be applied to higher orders of $\al_s$ expansion. But
to higher order in Feynman gauge, one has to deal with the $k_\perp$ expansion not only for fermion propagator, but also for gluon propagator and three-gluon vertex,
gluon-ghost vertex. The complicated color algebra makes the analysis
difficult. It is very likely that the use of Ward identities can simplify the
analysis and give a very general conclusion. Moreover, to higher order, collinear
and soft divergences will appear. For a complete analysis,
one has to provide a systematic subtraction scheme to get finite hard coefficients.
These issues are beyond the scope of this paper. Works on these aspects are ongoing.
When this work is finished, it is found that in\cite{Xing:2019ovj}, a similar formula eq.(35) is obtained with a different method. Ward identity is used in \cite{Xing:2019ovj} before pole condition is taken. The derivation and conclusion
there seems more general than ours. But as we known, the Ward identity
$\langle M|\partial_\mu G^\mu|N\rangle=0$ holds for complete physical amplitude,
which contains compact diagrams in Fig.\ref{fig:contact}. The treatment of these
contact diagrams at twist-3 is not transparent\cite{Boer:2001tx}, which may deserve
further study in the language of Ward identities. In our calculation, gluon field strength tensor simply appears as a consequence of eq.(\ref{eq:derivative_transverse_gluon}). The calculation seems
more transparent.

As a summary, we have shown how to write SGP contribution of SSA into a gauge
invariant form by including transverse gluon in the power expansion. The crucial
step is to include Fig.\ref{fig:Gperp} in the calculation. By including $G_\perp$
in the calculation, $G^+$ can be absorbed into gauge link and gluon field strength
tensor in a definite way. The resulting formula contains two parts:
one is related to $q_\partial(x)$, the other is to $T_F(x,x)$.
The coefficient of $q_\partial(x)$ can be obtained from calculating quark-photon
scattering amplitude with quark transverse momentum preserved. The coefficient of
$T_F$ can be obtained by calculating the subcross section for
$\ga^*+qg$ with initial gluon on-shell and transversely polarized. Since initial
gluon is on-shell and physically polarized, many tricks for on-shell
amplitudes can be applied, please see \cite{Elvang:2013cua} for example.
As a check, we calculate the hard coefficients in
SIDIS for pion production with Feynman and light-cone gauges, respectively. The results are the same as those given in
\cite{Eguchi:2006mc}. Generalizing this derivation to higher order of $\al_s$ expansion is interesting and can help us to understand the structure of twist-3 factorization formula.

\section*{Acknowledgements}
This work is supported by National Nature Science Foundation of China(NSFC) with
contract No.11605195.

\appendix
\section{Rules for special vertices}\label{sec:rules}
Some special vertices appear in our derivation due to expansion of $k_\perp$.
The rules for these vertices are
\begin{align}
\begin{matrix}
\includegraphics[width=0.2\textwidth]{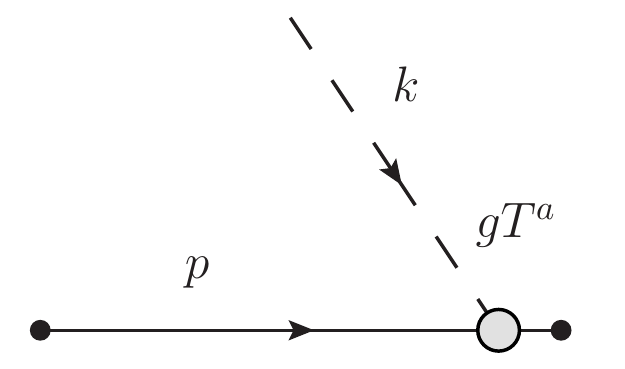}
\end{matrix}
=&gT^a\frac{i}{\s{p}+i\ep},\no
\begin{matrix}
\includegraphics[width=0.2\textwidth]{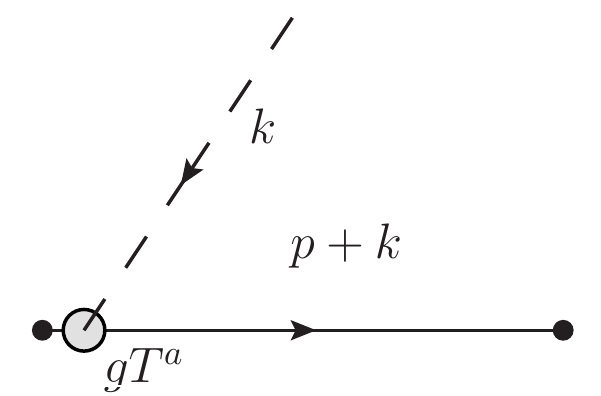}
\end{matrix}
=&\frac{i}{\s{p}+\s{k}+i\ep}gT^a,\no
\begin{matrix}
\includegraphics[width=0.2\textwidth]{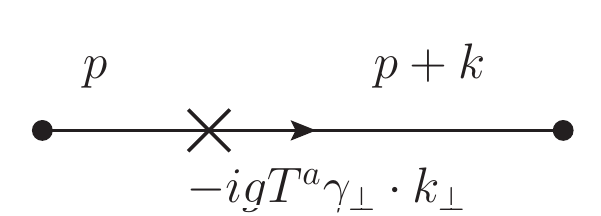}
\end{matrix}
=&\frac{i}{\s{p}+\s{k}_\perp+i\ep}(-igT^a\gamma_\perp\cdot k_\perp)
\frac{i}{\s{p}+i\ep}.
\end{align}

\section{Angular distributions}\label{sec:weights}
The projection tensors in \cite{Eguchi:2006mc} for hadronic tensor are
\begin{align}
&\tilde{\mathcal{V}}_1^{\mu\nu}=\frac{1}{2}(2T^\mu T^\nu+X^\mu X^\nu+Y^\mu Y^\nu),\
\tilde{\mathcal{V}}_2^{\mu\nu}=T^\mu T^\nu,\no
&\tilde{\mathcal{V}}_3^{\mu\nu}=\frac{1}{2}(T^\mu X^\nu+X^\mu T^\nu),\
\tilde{\mathcal{V}}_4^{\mu\nu}=\frac{1}{2}(X^\mu X^\nu-Y^\mu Y^\nu).
\end{align}
The momenta are
\begin{align}
T^\mu=&\frac{1}{Q}(q^\mu+2x_{BJ}P_A^\mu),\no
X^\mu=&\frac{1}{q_T}\left\{\frac{1}{z_f}P_H^\mu -q^\mu
-(1\textcolor{red}{+}\frac{q_T^2}{Q^2})x_{BJ}P_A^\mu\right\},\no
Y^\mu=&\ep^{\mu\nu\rho\sig}Z_\nu X_\rho T_\sig,\no
Z^\mu=&-q^\mu/Q,
\end{align}
which satisfy $T^2=-X^2=-Y^2=-Z^2=1$. Metric can be expressed as
\begin{align}
g^{\mu\nu}=T^\mu T^\nu -X^\mu X^\nu-Y^\mu Y^\nu-Z^\mu Z^\nu.
\end{align}
So, $Y^\mu Y^\nu$ can be eliminated.

With $p_A$ and $p_H$ chosen as two light-like reference vectors to define light-cone
coordinates, the momentum of virtual photon is decomposed as
\begin{align}
q^\mu=&\frac{q\cdot p_H}{p_A\cdot p_H}p_A^\mu
+\frac{q\cdot p_A}{p_A\cdot p_H}p_H^\mu +q_\perp^\mu.
\end{align}
Because
\begin{align}
q\cdot p_H=q^+ p_H^-=\frac{-Q^2+q_T^2}{2q^-}p_H^-=z_f\frac{-Q^2+q_T^2}{2},\
q_T=\sqrt{-q_\perp^2},
\end{align}
$X^\mu$ can be simplified as
\begin{align}
X^\mu=\frac{1}{q_T}\Big[-q_\perp^\mu-2\frac{q_T^2}{Q^2}x_B p_A^\mu\Big].
\end{align}
Interestingly, it has a longitudinal component.
The projection tensors are convenient to be rewritten as
\begin{align}
&\tilde{\mathcal{V}}_1^{\mu\nu}=\frac{1}{2}(3T^\mu T^\nu-Z^\mu Z^\nu-g^{\mu\nu}),\
\tilde{\mathcal{V}}_2^{\mu\nu}=T^\mu T^\nu,\no
&\tilde{\mathcal{V}}_3^{\mu\nu}=-\frac{1}{2}(T^\mu X^\nu+X^\mu T^\nu),\
\tilde{\mathcal{V}}_4^{\mu\nu}=\frac{1}{2}(2X^\mu X^\nu+g^{\mu\nu}-T^\mu T^\nu
+Z^\mu Z^\nu).
\end{align}

\section{Gauge dependence of $C_i^k$}\label{sec:c_i}
At this order, $O(\al)$, the gauge dependence only appears in the case with
quark fragmentation. In the following, $C_i^k$ are calculated in Feynman gauge
and light-cone gauge with $G^+=0$, respectively. The difference is defined
as
\begin{align}
\Delta C_i^k\equiv C_i^k|_{LC}-C_i^k|_{Feynman}.
\end{align}
The results are
\begin{align}
\Delta C_2^k=&\Delta C_4^k=0,\ \Delta C_3^k=\frac{1}{2}\Delta C_1^k,
\end{align}
and
\begin{align}
\Delta C_1^1=&2\frac{\al_sC_F}{\pi^2 Q^2}
\frac{x_B z_f \left(x_1 \left(-5 z z_f+3
   z_f^2+z^2\right)-x_B \left(-6 z z_f+6
   z_f^2+z^2\right)\right)}{x_1^2 z
   \left(x_1-x_B\right)
   \left(z-z_f\right){}^2},\no
\Delta C_1^2=&2\frac{\al_sC_F}{\pi^2 Q^2}
-\frac{2 \left(x_1-2 x_B\right) x_B
   z_f^2}{x_1^2 z \left(x_1-x_B\right)
   \left(z-z_f\right)},\no
\Delta C_1^3=&2\frac{\al_sC_F}{\pi^2 Q^2}
\frac{Q z_f \left(x_1 x_B \left(7 z-8
   z_f\right)-4 x_B^2 \left(z-2
   z_f\right)+x_1^2 \left(z_f-2
   z\right)\right)}{2 x_1^2 z
   \left(x_1-x_B\right) \left(z-z_f\right)
   q_T},\no
\Delta C_1^4=&2\frac{\al_sC_F}{\pi^2 Q^2}
\frac{x_B z_f \left(2 x_B z_f+x_1
   \left(z-z_f\right)\right)}{x_1^2 z
   \left(x_1-x_B\right) \left(z-z_f\right)}.
\end{align}

\bibliography{ref2}

\end{document}